\documentclass[12pt, draftclsnofoot, onecolumn]{IEEEtran}

\usepackage{color}
\usepackage{cite}
\usepackage{url}
\ifCLASSINFOpdf
\else
\fi
\usepackage{caption}
\usepackage{amsfonts,amssymb}
\usepackage{graphicx}
\usepackage{epstopdf}
\usepackage{mathrsfs}
\usepackage{amsmath}
\usepackage{latexsym}
\usepackage{booktabs}
\usepackage{subfigure}
\usepackage{algorithm, algorithmic}
\usepackage{bbm}
\usepackage[amsmath,thmmarks]{ntheorem}
\usepackage{bm}
\usepackage{extpfeil}

\DeclareCaptionFont{blue}{\color{blue}}

\begin{document}
\title{\LARGE Meta-Reinforcement Learning for Reliable Communication in THz/VLC Wireless VR Networks}

\author{\normalsize{{Yining Wang, \emph{Student Member, IEEE},  {Mingzhe Chen,} \emph{Member,~IEEE}, \\
{Zhaohui Yang,} \emph{Member,~IEEE}, {Walid Saad,} \emph{Fellow,~IEEE}, {Tao Luo,} \emph{Senior Member,~IEEE}, \\
{Shuguang Cui,} \emph{Fellow,~IEEE}, and {H. Vincent Poor,} \emph{Life Fellow,~IEEE}}}\vspace*{-3em}\\
\thanks{Y. Wang and T. Luo are with the Beijing Laboratory of Advanced Information Network, Beijing University of Posts and Telecommunications, Beijing, 100876, China, (e-mail: \protect\url{wyy0206@bupt.edu.cn}; \protect\url{tluo@bupt.edu.cn}).}
\thanks{M. Chen and H. V. Poor are with the Department of Electrical and Computer Engineering, Princeton University, Princeton, NJ, 08544, USA, (e-mail: \protect\url{mingzhec@princeton.edu}; \protect\url{poor@princeton.edu}).}
\thanks{Z. Yang is with the Department of Electronic and Electrical Engineering, University College London, WC1E 6BT London, UK, (e-mail: \protect\url{zhaohui.yang@ucl.ac.uk}).}
\thanks{W. Saad is with the Wireless@VT, Bradley Department of Electrical and Computer Engineering, Virginia Tech, Arlington, VA, 22203, USA, (e-mail: \protect\url{walids@vt.edu}).}
\thanks{S. Cui is currently with the School of Science and Engineering (SSE) and Future Network of Intelligence Institute (FNii), the Chinese University of Hong Kong, and Shenzhen Research Institute of Big Data, Shenzhen, China, 518172; he is also affiliated with Peng Cheng Laboratory, Shenzhen, China, 518066 (e-mail: \protect\url{shuguangcui@cuhk.edu.cn}).}
\thanks{This work was supported in part by the National Natural Science Foundation of China under Grant 62171047, in part by the US National Science Foundation under Grant CNS-1909372, in part by the National Key R\&D Program of China with grant No. 2018YFB1800800, in part by the Basic Research Project No. HZQB-KCZYZ-2021067 of Hetao Shenzhen-HK S\&T Cooperation Zone, in part by Shenzhen Outstanding Talents Training Fund 202002, in part by Guangdong Research Projects No. 2017ZT07X152 and No. 2019CX01X104, and in part by BUPT Excellent Ph.D. Students Foundation (CX2020210).}
\thanks{A preliminary version of this work \cite{wyy2020ICC} is published in the Proceedings of the 2021 IEEE International Conference on Communications.}
}

\maketitle


\begin{abstract}
In this paper, the problem of enhancing the quality of virtual reality (VR) services is studied for an indoor terahertz (THz)/visible light communication (VLC) wireless network. 
In the studied model, small base stations (SBSs) transmit high-quality VR images to VR users over THz bands and light-emitting diodes (LEDs) provide accurate indoor positioning services for them using VLC.
Here, VR users move in real time and their movement patterns change over time according to their applications, where both THz and VLC links can be blocked by the bodies of VR users.
To control the energy consumption of the studied THz/VLC wireless VR network, VLC access points (VAPs) must be selectively turned on so as to ensure accurate and extensive positioning for VR users.
Based on the user positions, each SBS must generate corresponding VR images and establish THz links without body blockage to transmit the VR content.
The problem is formulated as an optimization problem whose goal is to maximize the average number of successfully served VR users by selecting the appropriate VAPs to be turned on and controlling the user association with SBSs.
To solve this problem, a policy gradient-based reinforcement learning (RL) algorithm that adopts a meta-learning approach is proposed.
The proposed meta policy gradient (MPG) algorithm enables the trained policy to quickly adapt to new user movement patterns.
In order to solve the problem of maximizing the average number of successfully served users for VR scenarios with a large number of users, a dual method based MPG algorithm (D-MPG) with a low complexity is proposed.
Simulation results demonstrate that, compared to a baseline trust region policy optimization algorithm (TRPO), the proposed MPG and D-MPG algorithms yield up to 26.8\% and 21.9\% improvement in the average number of successfully served users as well as 81.2\% and 87.5\% gains in the convergence speed, respectively. 
\end{abstract}

\begin{IEEEkeywords} 
Wireless virtual reality, terahertz (THz), visible light communications (VLC), indoor positioning, meta-learning, reinforcement learning (RL), reliability.
\end{IEEEkeywords}

%{\renewcommand{\thefootnote}{\fnsymbol{footnote}}
%\footnotetext{A preliminary version of this work was submitted to the IEEE ICC 2021 \cite{wyy2020ICC}.}}

% \fi
\IEEEpeerreviewmaketitle

\vspace{-0.4cm}
\section{Introduction}
\vspace{-0.1cm}
Deploying virtual reality (VR) applications over wireless networks provides new opportunities for VR to offer seamless user experience \cite{VRsurvey_Walid}.
However, the scarce bandwidth of sub-6 GHz limits the ability of wireless networks to satisfy the stringent quality-of-service (QoS) requirements of VR applications in terms of delivering high data rates, low latency, and high reliability.
A promising solution is to integrate VR services over high frequency bands with abundant bandwidth, such as terahertz (THz) and millimeter wave (mmWave) frequencies.
Currently, 5G supports millimeter wave (mmWave) frequency bands to provide basic wireless VR services.
However, as discussed by industry in \cite{Huawei}, in order to support ultimate VR services that must integrate vision with perception, an uncompressed bit rate of up to 2 Tbit/s is strictly required.
Hence, it is necessary to study the use of frequency bands beyond mmWave for future wireless networks.
In addition, although the use of beamforming enables mmWave beams to focus on a small area, interference between neighboring users is still difficult to control in a dense room.
Therefore, THz frequencies are viewed as a natural candidate to provide unprecedentedly high data rate for VR content transmission due to the large available bandwidth.
Moreover, THz bands can achieve very narrow pencil beamforming (narrower than mmWave) that spatially aligns narrow THz beams to VR users and, hence significantly reducing the interference \cite{THz_zhanghaijun}.
However, THz frequencies are highly prone to blockage and their transmission distance is short \cite{THzVR_Walid}.
In indoor VR scenarios, although short distances enable high-rate VR image transmission at THz frequencies, the mobile users' bodies may lead to dynamic blockages over the THz links, thus negatively affecting the immersive VR experience.
In addition, to ensure a seamless interaction between the users and the virtual world, it is necessary to accurately locate VR users in real time for VR image generation and transmission.
Therefore, deploying THz-enabled wireless networks to offer high-reliability VR services faces many challenges such as user positioning, reduction of link blockage, user association, and reliability assurance.

Recently, several works such as in \cite{VR_JSAC_pengmugen, VR_TCOM, Yansha_VR, VR_Bennis, Network_Bennis, TM_mmWave, THzVR_Walid, THzVR_Walid_ICC} studied a number of problems related to wireless VR networks.
In \cite{VR_JSAC_pengmugen}, the authors studied the use of both edge fog computing and caching to satisfy the low latency requirement of VR users.
The authors in \cite{VR_TCOM} proposed a novel mobile edge computing-based mobile VR delivery framework that can cache the field of views (FOVs) of ${360^ \circ}$ VR images.
The work in \cite{Yansha_VR} studied the problem of resource management in wireless VR networks to minimize the VR interaction latency.
However, the works in \cite{VR_JSAC_pengmugen, VR_TCOM, Yansha_VR} sacrificed the quality of delivered VR videos (e.g., by reducing the resolution of VR videos or only displaying the FOV of ${360^ \circ}$ VR images) to meet the low latency constraints.
This challenge can be addressed by using high frequency bands (e.g., mmWave and THz) with abundant bandwidth to transmit high-quality VR images.
The authors in \cite{VR_Bennis} investigated the use of the mmWave bands to maximize the quality of the delivered video chunks in a wireless VR network.
In \cite{Network_Bennis}, the authors introduced a multi-connectivity (MC)-enabled mmWave network for providing low-latency VR services.
The work in \cite{TM_mmWave} studied the use of mmWave bands to meet the high bandwidth requirements of panoramic VR video streaming.
However, the works in \cite{VR_Bennis, Network_Bennis,TM_mmWave} did not study how to use mmWave and high frequency bands to provide reliable VR services in a dense VR scenario.
In \cite{THzVR_Walid}, the authors studied the use of THz bands to provide VR services in a dense VR network.
%verified that using THz bands, a successful transmission probability of 96\% (with the delay threshold of 20 ms) can be achieved even in the dense VR environment.
The authors in \cite{THzVR_Walid_ICC} studied the use of THz-based reconfigurable intelligent surfaces (RISs) to serve VR users in a wireless network.
However, the works in \cite{THzVR_Walid} and \cite{VR_Bennis, Network_Bennis, TM_mmWave, THzVR_Walid_ICC} did not consider the mobility of users that can significantly affect VR network performance, particularly for THz-enabled wireless VR networks whose transmission links can be blocked by mobile users.
Moreover, all of the existing works in \cite{VR_JSAC_pengmugen, VR_TCOM, Yansha_VR, VR_Bennis, Network_Bennis, TM_mmWave, THzVR_Walid, THzVR_Walid_ICC} ignored the requirement of accurate user localization that is needed to generate users' VR images.
Therefore, in a THz-enabled VR system, it is necessary to consider the time-varying user positions that are used to generate VR images and avoid dynamic blockages of THz links.

A number of existing works such as in \cite{VR_TWC_chenmingzhe, Ultrawideband, Laser} studied the problem of positioning applied in a VR system.
%In \cite{VR_TWC_chenmingzhe} and \cite{letter_walid}
In \cite{VR_TWC_chenmingzhe}, the authors used machine learning (ML) algorithms to predict the locations and orientation of VR users.
However, the position prediction accuracy of ML algorithms depends on the training data and cannot adapt to different users' movement patterns.
The authors in \cite{Ultrawideband} studied the use of ultrawideband signals and ultrasonic waves to achieve decimeter-level VR user positioning, respectively.
The work in \cite{Laser} proposed a mobile laser scanning (MLS) positioning system for indoor VR applications.
Although the positioning accuracy of an MLS system can reach the centimeter-level accuracy, such a laser system is expensive.
Moreover, the existing works in \cite{Ultrawideband} and \cite{Laser} require equipping VR systems with additional positioning devices, thus increasing energy consumption and deployment costs.
The work in \cite{THz_positioning_JSAC} showed that THz has the potential for indoor positioning.
However, since THz bands require very narrow pencil beamforming in dense indoor VR scenarios, one can only passively adjust the beam direction or user association after the user moves, which can detach the users from their virtual world.
Visible light communication (VLC) based on light-emitting diodes (LEDs) can provide an alternative and accurate positioning service \cite{Positioning_Survey}.
In \cite{VLC_accuracy, VLC_accuracy2, VLC_accuracy6, Ismail}, the authors proved that using three LEDs that are in the line of sight (LoS) of the receiver can provide a centimeter-level three-dimensional (3-D) position.
However, none of these works in \cite{VLC_accuracy, VLC_accuracy2, VLC_accuracy6, Ismail} considered the dynamic selection of LEDs according to the user mobility so as to provide inclusive positioning services while ensuring acceptable brightness in a multi-user VR scenario.
To this end, we propose to use a THz/VLC-enabled wireless VR network that jointly considers the VLC access points (VAPs) selection and user association in order to provide reliable positioning and high data rate VR content transmission services for VR users.

The main contribution of this work is, thus, a novel framework that jointly uses VLC and THz to service VR users.
In particular, we study a dynamic THz/VLC-enabled VR network that can accurately locate VR users in real time using VLC and build THz links to transmit high-quality VR images based on the users' positions.
In the studied network, only a subset of the VAPs can be turned on to locate VR users due to the users' limited tolerance for brightness.
Based on the obtained user positions, each small base station (SBS) must determine the user association to generate corresponding VR images and build THz links to avoid blockages caused by the user bodies.
The problem is formulated as a reliability maximization problem that jointly considers the VAP selection, user association with THz SBSs, and time varying users' movement patterns. 
The \emph{reliability} of VR networks is defined as the average number of successfully served VR users.
To solve this problem, we propose a meta-policy gradient (MPG) algorithm to find the locally optimal policy for VAP selection and user association.
Compared to traditional reinforcement learning (RL) algorithms that can only be trained for a fixed environment in which each user has a fixed movement pattern, the proposed algorithm enables the trained policy to quickly adapt to new users' movement patterns.
To reduce the computational complexity of the MPG algorithm, we propose a dual method based MPG algorithm that uses dual method to assist the MPG algorithm to determine user association based on the selected VAPs.
Simulation results show that, compared to a baseline trust region policy optimization algorithm (TRPO), the proposed MPG algorithm and the dual method based MPG algorithm yield a performance improvement of about 26.8\% and 21.9\% in terms of the average number of successfully served users as well as about 81.2\% and 87.5\% gains in the convergence speed, respectively. 
Simulation results also show that the proposed dual method based MPG algorithm achieves up to 88.7\% reduction in the training time compared to the MPG algorithm.
To the best of our knowledge, \emph{this paper is the first to study the joint use of THz and VLC for reliability maximization while considering dynamic VR users' movement patterns}.

The rest of this paper is organized as follows. 
The system model and the problem formulation are described in Section~\ref{sec:2}. 
The use of MPG algorithm for VAP selection and user association is introduced in Section~\ref{sec:3}. 
The dual method based MPG algorithm is presented in Section~\ref{sec:4}.
In Section~\ref{sec:5}, the numerical results are presented and discussed. 
Finally, conclusions are drawn in Section~\ref{sec:6}.

\vspace{-0.4cm}
\section{System Model and Problem Formulation}
\label{sec:2}
\begin{figure}[t]
\centering
\setlength{\abovecaptionskip}{-0cm}
\setlength{\belowcaptionskip}{-0.8cm}
\includegraphics[width=8.5cm]{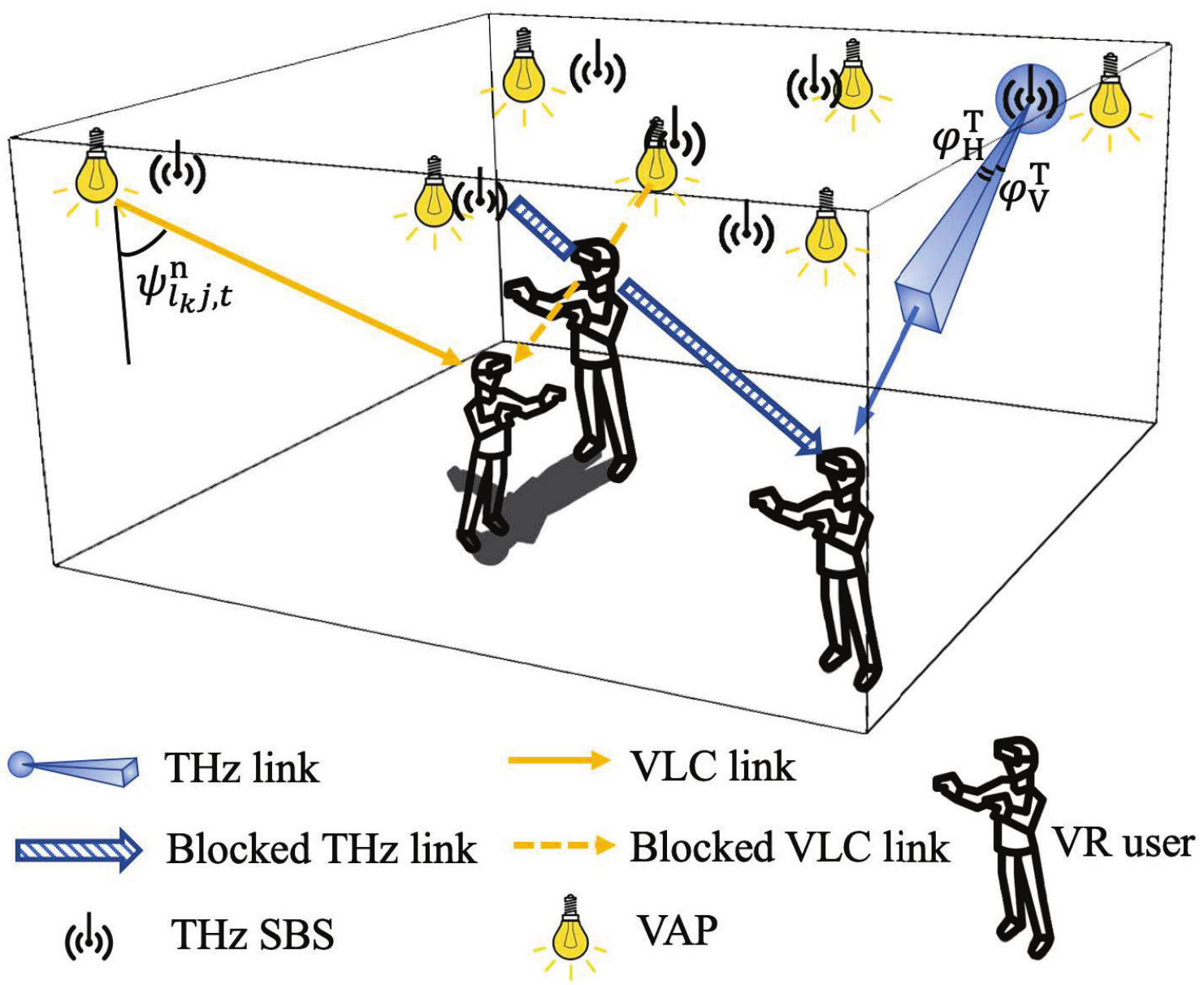}
%\captionsetup{labelfont={blue},textfont={blue}}
\caption{Illustration of the considered THz/VLC-enabled wireless VR network.}
\label{fig1}
\end{figure}

\vspace{-0.1cm}
Consider an indoor wireless network that consists of a set $\cal{B}$ of $\emph{B}$ SBSs and a set $\cal{V}$ of $\emph{V}$ VAPs. 
All the VAPs and SBSs are managed by a central controller.
The SBSs are evenly distributed in an indoor area ${\cal G}$ to serve a set $\cal{U}$ of $\emph{U}$ VR users over THz frequencies, as shown in Fig. \ref{fig1}. 
In the studied model, accurate locations of the users are required by the SBSs so as to build LoS THz links and generate the VR images requested by users \cite{VR_TCOM}.
Each VAP provides accurate indoor positioning and tracking services for VR users using VLC.
% that consists of a set of light emitting diodes (LEDs)
Here, we consider dual-mode user equipments (UEs) that are able to access both THz and VLC bands.
In the studied multi-user VR network, at each time slot $t$, each SBS can only serve one user with a narrow beam while each VAP can locate all the users that are not blocked in its FOV.
To control the system energy consumption, the central controller selects a group of VAPs at the beginning of each time slot to locate VR users.
Here, not all users can be accurately localized due to the user body blockage over the VLC links \cite{Yangyang}.
Hence, based on the obtained user positions, the central controller determines the SBSs associated with the successfully localized users, and then SBSs transmit the corresponding VR images to those users using wireless THz links.
In our model, each time period $n$ consists of $T$ time slots.
A successful transmission implies that the request of a given VR user is successfully completed within a time period.
% we assume that the VR users do not move until the VAPs complete the localization

\vspace{-0.2cm}
\subsection{User Blockage Model}
\vspace{-0.1cm}
In the studied model, the LoS links (VLC or THz links) between user $j$ and a transmitter (a VAP or an SBS) can be blocked by other VR users’ bodies \cite{THzVR_Walid}.
%due to the inability to penetrate solid objects for VLC and THz communications.
%Since the susceptibility of VLC and THz to blockage jeopardize the reliability of the considered VR network, we must analyze the blockage based on the positions of users, VAPs, and SBSs.
For a given a user $j$ located at $\bm{v}_{j,t}^n=(x_{j,t}^n,y_{j,t}^n,z_{j,t}^n)$ at time slot $t$ in time period $n$ and a transmitter $k$ located at $(x_k,y_k,Z)$, we define a binary variable $b_{kj,t}^n$ that indicates whether LoS links exist between user $j$ and transmitter $k$, as follows: 
\vspace{-0.2cm}
\begin{equation}
\label{blockage}
b_{kj,t}^n\!=\!\mathbbm{1}_{\!\left\{\!\!\left\{\!\gamma(x_k,y_k,Z)+(1\!-\!\gamma)(x_{j,t}^n,y_{j,t}^n,z_{j,t}^n)|0<\gamma<1\!\right\}\cap\!\!\!\!\bigcup\limits_{m \ne j\in {\cal{U}}}\!\!\!\!\!{\cal{D}}_{m,t}^n=\emptyset\!\right\}},
\end{equation}
where $\left\{\!\gamma(x_k,y_k,Z)+(1\!-\!\gamma)(x_{j,t}^n,y_{j,t}^n,z_{j,t}^n)|0<\gamma<1\!\right\}$ is the set of all points in the LoS transmission link between transmitter $k$ and user $j$, ${\cal{D}}_{m,t}^n=\{(x_{m,t}^n,y_{m,t}^n,z)|0\leqslant z \leqslant z_{m,t}^n\}$ is the space occupied by the body of user $m$ at time slot $t$ in time period $n$ and $\mathbbm{1}_{\{x\}}=1$ as $x$ is true, $\mathbbm{1}_{\{x\}}=0$, otherwise.
Equation (\ref{blockage}) indicates that the LoS link between transmitter $k$ and user $j$ at time slot $t$ exists only if none of the other users ($m\ne j\in {\cal{U}}$) blocks the transmission link, as shown in Fig. \ref{fig1}.
In (\ref{blockage}), $b_{kj,t}^n=0$ implies that the link between transmitter $k$ and user $j$ is blocked at time slot $t$ in period $n$; otherwise, we have $b_{kj,t}^n=1$.
Here, we assume that the positions of the VR users remain unchanged during each time slot $t$.
%Here, we assume that the VR users are static during each time slot $t$.

\vspace{-0.3cm}
\subsection{VLC Indoor Positioning}
\vspace{-0.1cm}
We assume that the three-dimensional (3D) location $\bm{v}_{j,t}^n=(x_{j,t}^n,y_{j,t}^n,z_{j,t}^n)$ of each user $j$ is determined by three VAPs from three different orientations \cite{Positioning_Survey}, where $x_{j,t}^n$ and $y_{j,t}^n$ are the coordinates of user $j$ in the room, and $z_{j,t}^n$ is the height of user $j$.
Here, we consider the use of only three VAPs to localize each user since using more VAPs may increases the complexity of the positioning algorithm \cite{VLC_accuracy} and the energy consumption for user localization.

At each time slot $t$ in time period $n$, a set ${\cal{L}}_t^n=\{l_1, l_2, l_3\}$ of three VAPs is turned on to broadcast their location information to users.
We assume that no optical filter and concentrator will be used.
The LoS channel gain of the VLC link between VAP $l_k$ and user $j$ will be given by \cite{UAV_VLC}
\vspace{-0.2cm}
\begin{equation}
\vspace{-0.1cm}
\label{VLC_LoS}
G_{{l_k}j,t}^n= \left\{ {\begin{array}{*{20}{l}}
{A_{{l_k}j,t}^n, \;0 \leqslant \psi _{{l_k}j,t}^n \leqslant \Psi_{\frac{1}{2}}\; {\rm{and}}\;b_{{l_k}j,t}^n=1,}\\
{\;\;\;\;0,\;\;\;\; {\rm{others}},}
\end{array}} \right.
\end{equation}
where $A_{{l_k}j,t}^n=\frac{{(m + 1)\rho }}{{2\pi {{(d_{{l_k}j,t}^n)}^2}}}{\cos ^m}(\phi _{{l_k}j,t}^n){\cos ^M}(\psi _{{l_k}j,t}^n)$ with $\rho$ being the detector area, $d_{{l_k}j,t}^n$ being the distance between VAP $l_k$ and user $j$, $\Psi_{\frac{1}{2}}$ being the receiver FOV semi-angle, $\phi _{{l_k}j,t}^n$ and $\psi _{{l_k}j,t}^n$ being the angle of irradiance and incidence, respectively, $m=- \ln 2/\ln (\cos(\Phi_{\frac{1}{2}}))$ and $M=- \ln 2/\ln (\cos(\Psi_{\frac{1}{2}}))$ being Lambertian parameters that depend on the half-power angle $\Phi_{\frac{1}{2}}$ of VAP and the receiver FOV semi-angle $\Psi_{\frac{1}{2}}$, respectively.
From (\ref{VLC_LoS}), we see that user $j$ can receive the location information sent by VAP $l_k$ at time slot $t$ only when the following conditions are satisfied: a) VAP $l_k$ is in the FOV of user $j$, and b) the VLC link between VAP $l_k$ and user $j$ is LoS (i.e. $b_{{l_k}j,t}^n=1$). 
When user $j$ receives the location information of three VAPs, it can accurately calculate its location.
Here, we ignore the centimeter-level positioning error since the VLC based localization is accurate enough for building THz links, generating VR images, and user blockage analysis \cite{VLC_accuracy, VLC_accuracy2, VLC_accuracy6}.
Then, the set of VAPs available for providing the positioning service to user $j$ can be given by
\vspace{-0.1cm}
\begin{equation}
\vspace{-0.1cm}
\label{VAPs}
{\cal{L}}_{j,t}^n\!=\!\left\{l_k\left|0\!\leqslant\! {\psi}_{{l_k}j,t}^n\! \leqslant\! \Psi_{\frac{1}{2}},b_{l_kj,t}^n=1, l_k \!\in\! {\cal{L}}_t^n \right.\right\},
\end{equation}
where $\Psi_{\frac{1}{2}}$ is the receiver FOV semi-angle.

Based on three different incidence angles and the corresponding VAP locations, each user $j$ can calculate its own location $\bm{v}_{j,t}^n$ at time slot $t$ in period $n$ using a triangulation algorithm \cite{VLC_accuracy6}.
Then, the positioning state of user $j$ at time slot $t$ in period $n$ will be
\vspace{-0.1cm}
\begin{equation}
\vspace{-0.1cm}
\label{loc_state}
{p_{j,t}^n}({\cal{L}}_t^n) = \left\{ {\begin{array}{*{20}{l}}
{1, |{\cal{L}}_{j,t}^n| = 3,}\\
{0, |{\cal{L}}_{j,t}^n| < 3,}
\end{array}} \right.
\end{equation}
where $|{\cal{L}}_{j,t}^n|$ represents the number of VAPs that can serve user $j$.

Once the position of user $j$ is successfully calculated at time slot $t$ in period $n$ (i.e. $p_{j,t}^n({\cal{L}}_t^n)=1$), user $j$ transmits its own location to the central controller and requests the corresponding VR image.
Here, we do not consider the time that each user transmits its position information transmission to the central controller, since the location information of each user only consists of three scalars and since this location information is transmitted over THz frequencies that have abundant bandwidth. 
Based on the obtained user positions, the central controller can determine the user-SBS association and, then, the SBSs can generate corresponding VR images and serve the associated users over THz band.
%As done in \cite{}, we assume that the position of each user during the VR service follows the Markov chain which is given as:
%\begin{equation}
%\label{Markov}
%Pr\!\left({\bm{c}}_{j,t+1}|{\bm{c}}_{j,t},{\bm{c}}_{j,t-1},\!\cdots\!,{\bm{c}}_{j,1}\right)=Pr\!\left({\bm{c}}_{j,t+1}|{\bm{c}}_{j,t}\right).
%\end{equation}
%From (\ref{Markov}), we can see that the location distribution of each user $j$ at next time slot depends on the location at current location.

\vspace{-0.3cm}
\subsection{Transmission Model}
\vspace{-0.1cm}
We assume that a time division multiple access (TDMA) technique is adopted for each SBS.
Due to the extremely narrow pencil beamforming (narrower than mmWave) for THz \cite{THz_zhanghaijun}, we assume that each user can only be associated with one SBS and each SBS can only serve one user at each time slot.
In time period $n$, let $u_{ij,t}^n \in \{0,1\}$ be the index of the link between SBS $i$ and user $j$ at time slot $t$, i.e., $u_{ij,t}^n=1$ implies that user $j$ is associated with SBS $i$; otherwise, we have $u_{ij,t}^n=0$.
Then, we have
\begin{equation}
\vspace{-0.1cm}
\label{usre_asso}
0 \leqslant \sum\limits_{i=1}^B u_{ij,t}^n \leqslant 1, \forall j \in {\cal U}, \;\;0 \leqslant \sum\limits_{j=1}^U u_{ij,t}^n \leqslant 1, \forall i \in {\cal B}.
\end{equation}
Since VR users move in real time, at different time periods, the VR users must be served by SBSs deployed in different locations to avoid blockages of the THz links and to meet the transmission delay constraints.
In the considered network, a time period consists of $T$ time slots.
Each user only needs to successfully receive the requested VR image within a single time slot in each time period to ensure an immersive VR experience.
Hence, the handovers between different SBSs can be completed during the time slots in which the SBSs do not transmit the VR images over THz links \cite{handover_JSAC}.
%handover_JSAC2, handover_TVT

In the studied model, we assume that each THz SBS and each user will be equipped with one directional antenna used to form a 3D THz beam.
One 3D THz beam is approximated by a 3D pyramidal-plus-sphere sectored antenna model \cite{3D_THz}.
In particular, the pyramidal zone accounts for the main lobe of the antenna beam while the sphere accounts for the side lobes of the antenna beam, as shown in Fig. \ref{fig1}.
At each time slot $t$, each SBS adjusts the direction of main lobe towards its associated user so as to guarantee beam alignment.
Hence, we only consider the transmit gain of the main lobe of each SBS, given by \cite{3D_THz}
\vspace{-0.2cm}
\begin{equation}
\vspace{-0.1cm}
\label{transmit_gain}
{G_{\rm{T}}} = \frac{4\pi}{(\iota_{\rm{T}}+1){\Omega_{\rm{T}}}},
\end{equation}
where $\iota_{\rm{T}}$ is the ratio of the power concentrated along the side lobes to the power concentrated along the main lobe of a transmit antenna and $\Omega_{\rm{T}}=4\arcsin \left(\tan \left(\frac{{\varphi _{\rm{H}}^{\rm{T}}}}{2}\right)\tan \left(\frac{{\varphi _{\rm{V}}^{\rm{T}}}}{2}\right)\right)$ with ${\varphi _{\rm{H}}^{\rm{T}}}$ and ${\varphi _{\rm{V}}^{\rm{T}}}$ being the horizontal and vertical beamwidths of the transmit antennas, respectively.
The receive gain of the directional antenna of each user will be
\vspace{-0.2cm}
\begin{equation}
\vspace{-0.1cm}
\label{receive_gain}
{G_{\rm{R}}} = \frac{4\pi}{(\iota_{\rm{R}}+1){\Omega_{\rm{R}}}},
\end{equation}
where $\iota_{\rm{R}}$ is the power ratio between the side lobes and the main lobe of a receive antenna and $\Omega_{\rm{R}}=4\arcsin \left(\tan \left(\frac{{\varphi _{\rm{H}}^{\rm{R}}}}{2}\right)\tan \left(\frac{{\varphi _{\rm{V}}^{\rm{R}}}}{2}\right)\right)$ with ${\varphi _{\rm{H}}^{\rm{R}}}$ and ${\varphi _{\rm{V}}^{\rm{R}}}$ being the horizontal and vertical beamwidths of the receive antennas, respectively.
Due to the abundant bandwidth at THz frequencies, we allocate orthogonal THz bands for each SBS to ensure that no inter-cell interference occurs.
In the studied model, since the transmission delay is limited within the time duration $\Delta{t}$ of a time slot and $\Delta{t}$ is in milliseconds, we assume that the users are static during transmission. 
Hence, we can reasonably ignore the Doppler effect in the considered transmission model.
At time slot $t$ in period $n$, given an SBS $ i \in {\cal B}$ located at $({x_i},{y_i},Z)$ and its associated user $j \in {\cal U}$ located at $(x_{j,t}^n,y_{j,t}^n,z_{j,t}^n)$, the path loss of the THz link between SBS $i$ and user $j$ can be given by \cite{THzVR_Walid}
\vspace{-0.1cm}
\begin{equation}
\vspace{-0.1cm}
\label{path_loss}
g_{ij,t}^n = \left\{ {\begin{array}{*{20}{l}}
{{\left(\frac{c}{{4\pi fr_{ij,t}^n}}\right)^2}{{\delta (r_{ij,t}^n)}}, \;b_{ij,t}^n=1,}\\
{\;\;\;\;\;\;\;\;\;\;\;\;0,\;\;\;\;\;\;\;\;\;\;\;\;b_{ij,t}^n=0,}
\end{array}} \right.
\end{equation}
where $r_{ij,t}^n\!=\!\sqrt {(x_i-x_{j,t}^n)^2+(y_i-y_{j,t}^n)^2+(Z-z_{j,t}^n)^2}$ is the distance between SBS $i$ and user $j$, $c$ is the speed of light, $f$ is the operating frequency, and $\delta (r_{ij,t}^n) \approx {e^{( - K(f)r_{ij,t}^n)}}$ represents the transmittance of the medium following the Beer-Lambert law with $K(f)$ being the overall absorption coefficient of the medium at THz frequency $f$\cite{THz_VTC}.
In (\ref{path_loss}), $b_{ij,t}^n$ is a binary variable that indicates whether a LoS link exists between SBS $i$ and user $j$ at time slot $t$ in period $n$. 
Given the location of user $j$, $b_{ij,t}^n$ can be obtained using (\ref{blockage}).
The total noise power at each UE $j$ that is generated by thermal agitation of electrons and molecular absorption is \cite{THz_VTC}
\vspace{-0.1cm}
\begin{equation}
\vspace{-0.1cm}
\label{noise}
I_{j,t}^n=I_0+\sum\limits_{l \in {\cal B}} {P\left(\frac{c}{{4\pi fr_{lj,t}^n}}\right)^2}{{\left(1-\delta (r_{lj,t}^n)\right)}},
\end{equation}
where $P$ is the transmit power of each SBS, $I_0={K_B}T_e$ represents the Johnson-Nyquist noise generated by thermal agitation of electrons in conductors with $K_B$ and $T_e$ being Boltzmann constant and the temperature in Kelvin, respectively, and $\sum\limits_{l \in {\cal B}} {P\left(\frac{c}{{4\pi fr_{lj,t}^n}}\right)^2}{{\left(1-\delta (r_{lj,t}^n)\right)}}$ is the sum of molecular absorption noise caused by the transmit power of any SBS $l \in {\cal B}$.
The data rate of VR image transmission from SBS $i$ to its associated user $j$ at time slot $t$ in period $n$ will then be
\vspace{-0.2cm}
\begin{equation}
\vspace{-0.1cm}
\label{rate}
C_{ij,t}^n(u_{ij,t}^n)=u_{ij,t}^nW{\log _2}\left(1 + \frac{{Pg_{ij,t}^n{G_{\rm{T}}}{G_{\rm{R}}}}}{{I_{j,t}^n}}\right).
\end{equation}
where $W$ is the bandwidth of the THz band.

%In the studied system, once the location ${\bm{c}}_{j,t}$ of user $j$ at time slot $t$ is obtained, the SBSs first determine the user association, and, then each SBS transmits the corresponding VR image to its associated user using THz band.
Given the data size $S$ of the VR image requested by user $j$ at time slot $t$ in period $n$, the transmission delay will be
\vspace{-0.1cm}
\begin{equation}
\vspace{-0.1cm}
\label{tran_delay}
d_{j,t}^n(\bm{u}_{j,t}^n)=\frac{S}{\sum\limits_{i=1}^B C_{ij,t}^n(u_{ij,t}^n)},
\end{equation}
%\begin{equation}
%\label{tran_delay2}
%d_{j,t}({\bm{c}}_{j,t}, \bm{u}_{j,t})={\rm{min}}\left\{\frac{S}{C_{ij,t}}\right\},\forall i \in \cal{B},
%\end{equation}
where $\bm{u}_{j,t}^n=[u_{1j,t}^n,u_{2j,t}^n,\cdots, u_{Bj,t}^n]$.
Note that the data size $S$ of a VR image only depends on the image resolution which remains unchanged during service.
Since the user position will change at the next time slot, the VR image requested by user $j$ can be successfully transmitted only when the transmission delay is within the time duration $\Delta t$ of a time slot $t$.
Then, in time period $n$, the transmission state of user $j$ at time slot $t$ can be expressed as
\vspace{-0.1cm}
\begin{equation}
\vspace{-0.1cm}
\label{tran_state}
{h_{j,t}^n}(\bm{u}_{j,t}^n) = \left\{ {\begin{array}{*{20}{l}}
{1, \;d_{j,t}^n(\bm{u}_{j,t}^n)\leqslant \Delta t,}\\
{0, \;{\rm{otherwise}}.}
\end{array}} \right.
\end{equation}
From (\ref{tran_state}), we can see that, whether the requested VR image of user $j$ is successfully transmitted at time slot $t$ or not depends on the user's locations, user association, and blockages between SBS $i$ and user $j$.  

\vspace{-0.4cm}
\subsection{Reliability Model}
\vspace{-0.1cm}
As mentioned earlier, in our model, the reliability of the THz/VLC-enabled wireless VR network refers to the average number of successfully served VR users.
At each time slot $t$, a successfully served user $j$ must satisfy two conditions: a) user $j$ is successfully localized and b) the VR image requested by user $j$ is transmitted within $\Delta t$.
In order to enable a seamless and immersive wireless VR experience, we assume that the waiting delay is limited to a time period that consists of $T$ time slots.
In other words, each user should be successfully served at least once in a time period.
Therefore, in time period $n$, the service state of user $j$ until time slot $t$ based on the selected ${\cal{L}}_{t}^n$ and ${\bm{u}}_{j,t}^n$ will be
\vspace{-0.2cm}
\begin{equation}
\label{service_state}
{w_{j,t}^n}({\cal{L}}_{t}^n,{\bm{u}}_{j,t}^n)=\left({p_{j,t}^n}({\cal{L}}_{t}^n){h_{j,t}^n}(\bm{u}_{j,t}^n)\right)\vee {w_{j,t-1}^n}({\cal{L}}_{t-1}^n,{\bm{u}}_{j,t-1}^n).
\end{equation}
where $t=2,3,\cdots,T$ and $\vee$ represents the logical ``or" operation.
The newly served users at time slot $t$ will be
\vspace{-0.2cm}
\begin{equation}
\label{user_served}
{\cal{W}}_t^n({\cal{L}}_{t}^n\!,\!{\bm{u}}_{j,t}^n)\!=\!\{ j|{w_{j,t}^n}({\cal{L}}_{t}^n\!,\!{\bm{u}}_{j,t}^n)\!=\!1,{w_{j,t-1}^n}({\cal{L}}_{t-1}^n\!,\!{\bm{u}}_{j,t-1}^n)\!=\!0\}.
\end{equation}
Then, the number of successfully served users in each time period $n$ can be given by
\vspace{-0.2cm}
\begin{equation}
\vspace{-0.1cm}
\label{reliability}
R^n({\cal{L}}_{:T}^n,{\bm{u}}_{j,:T}^n)=\sum\limits_{t = 1}^T {\left|{\cal{W}}_t^n({\cal{L}}_{t}^n\!,\!{\bm{u}}_{j,t}^n)\right|},
\end{equation}
where ${\cal{L}}_{:T}^n=\{{\cal{L}}_{1}^n,\cdots,{\cal{L}}_{T}^n\}$ and ${\bm{u}}_{j,:T}^n=\{{\bm{u}}_{j,1}^n,\cdots,{\bm{u}}_{j,T}^n\}$.
From (\ref{reliability})-(\ref{service_state}) we can see that, once a VR user is successfully served at least once in $T$ time slots, the seamless VR experience of this user can be guaranteed.
Therefore, the maximum waiting delay for each VR user is $T$ time slots.
%From (\ref{service_state}) we can see that, after a whole time period, the service state of VR users is determined by a trajectory $\xi=({\bm{s}_{1}^n}, {\bm{a}}_1^n, {\bm{s}_{2}^n}({\bm{a}}_1^n), {\bm{a}}_2^n, \cdots, {\bm{a}}_{T-1}^n,{\bm{s}_{T}^n}({\bm{a}}_{T-1}^n))$, where ${\bm{s}_{1}^n}=\{\bm{s}_{1,1}^n,\cdots,\bm{s}_{j,1}^n,\cdots,\bm{s}_{U,1}^n\}$ is the .

\vspace{-0.5cm}
\subsection{Problem Formulation}
\vspace{-0.1cm}
Given the defined system model, our goal is to effectively select the subset of VAPs to provide accurate positioning services and, then, determine the user-SBS association based on the obtained user positions so as to maximize the reliability of the studied VR network.
Then, the reliability maximization problem is formulated as follows:
\begin{subequations}
\label{optimization}
\setlength{\abovedisplayskip}{3pt}
\setlength{\belowdisplayskip}{2pt}
\begin{align}\tag{\theequation}
&{\mathop {\max }\limits_{{\cal{L}}_t^n,{\bm{u}}_{j,t}^n} \;\;\sum\limits_{n=1}^N\frac{R^n({\cal{L}}_{:T}^n,{\bm{u}}_{j,:T}^n)}{N},}\\
&{\;\;\;{\rm{s}}.{\rm{t}}.\;\;\;\;|{\cal{L}}_t^n|=3,}\\
&{\;\;\;\;\;\;\;\;\;\;\;\;0 \leqslant \sum\limits_{i=1}^B u_{ij,t}^n \leqslant 1, \forall j \in {\cal U},}\\
&{\;\;\;\;\;\;\;\;\;\;\;\;0 \leqslant \sum\limits_{j=1}^U u_{ij,t}^n \leqslant 1, \forall i \in {\cal B},}\\
&{\;\;\;\;\;\;\;\;\;\;\;\;u_{ij,t}^n \in \{0,1\}, \forall i \in {\cal B},\forall j \in {\cal U},}
\end{align}
\end{subequations}
where $N$ is the total number of all time periods.
Constraint (\ref{optimization}a) captures the fact that only three VAPs are selected at each time slot to provide positioning service.
Constraints (\ref{optimization}b), (\ref{optimization}c), and (\ref{optimization}d) indicate that each user can only be associated with one SBS and each SBS can only serve one user at each time slot.
From (\ref{optimization}), we can see that the reliability depends on the selected VAPs and the user association with SBSs.
Meanwhile, the VAP selection and the user-SBS association depend on the positions of the VR users.
However, the users’ positions continuously change as time elapses.
Therefore, real-time user positions are needed by the central controller so as to generate corresponding VR images and build THz links without blockages.
Moreover, due to the time-varying nature of VR applications, the users' movement pattern varies over different time periods \cite{mobility_zhangrui}.
Here, we define a position transition matrix $\bm{M}^n$ as the users' movement pattern during time period $n$, in which each element $\bm{M}^n_{\bm{v}_{j,t}^n,\bm{v}_{j,t+1}^n}=P(\bm{v}_{j,t+1}^n|\bm{v}_{j,t}^n)$ is the probability of the user moving from $\bm{v}_{j,t}^n$ to $\bm{v}_{j,t+1}^n$.
Note that the studied THz/VLC-enabled VR network has no knowledge of the users' movement patterns.
Due to the non-convexity and the unpredictability of the users' movement patterns, (\ref{optimization}) cannot be solved by the traditional optimization algorithms, such as dynamic programming or nonlinear programming.
Moreover, traditional RL algorithms, such as Q-learning \cite{WSH_Qlearning} or deep Q-network \cite{THzVR_Walid_ICC}, can only solve optimization problems in static and known environments, and, thus, they are also not suitable to solve the problem in (\ref{optimization}).
Hence, we propose a RL algorithm based on a meta-learning framework to sensitively adapt to dynamic users' movement patterns so as to determine the VAP selection and the user association in advance.
We next introduce a meta-reinforcement learning algorithm to proactively determine the VAP selection and the user association. 

\vspace{-0.3cm}
\section{Meta-Learning for VAP Selection and User Association}
\vspace{-0.1cm}
\label{sec:3}
Next, we introduce a policy gradient-based RL algorithm \cite{VPG} using meta-learning framework \cite{MAML}, called meta policy gradient (MPG), that can effectively solve problem (\ref{optimization}).
Traditional policy gradient algorithms can only determine the VAP selection and user association in a fixed environment (i.e., the fixed users' movement patterns).
Meta-learning is a novel learning approach that can integrate the prior reliability-enhancing experience with information collected from the new users' movement patterns, thus training a rapidly adaptive learning model.
Therefore, the proposed MPG can obtain the VAP selection and user association policies that can be quickly updated to adapt to new users' movement patterns using only a few further training steps.
Compared with the meta-trained value decomposition-based RL algorithm \cite{meta_Huye} that uses each agent's local observation of the environment to estimate the rewards resulting from the actions, the proposed MPG algorithm enables the agent to directly obtain the reward of a chosen action from the global environment.
Hence, the proposed MPG algorithm can effectively find a better action that results in a higher reliability compared to the RL algorithm in \cite{meta_Huye}.
%action that benefits for the reliability of the entire network.
The VAP selection aims to obtain the positions of as many users as possible under the limitation of energy consumption. 
Then, the user-SBS association is determined based on the user positions in a way to avoid blockages of THz links and meet the transmission delay constraints.
Next, we first introduce the components of the MPG algorithm for VAP selection and user association. 
Then, we explain the entire procedure of using our MPG algorithm to select VAPs and determine the user association with SBSs.

\vspace{-0.4cm}
\subsection{Components of MPG}
\vspace{-0.1cm}
An MPG algorithm consists of six components: a) agent, b) actions, c) states, d) policy, e) reward, and f) tasks, which are specified as follows: 
\begin{itemize}
\item \emph{Agent}: Our agent is a central controller that can obtain the user positions and simultaneously control the VAPs and the SBSs.
%\item \emph{Environment}: The time-varying positions of users are considered in our dynamic environment.
%The users are randomly distributed at the beginning and move based on a transition probabilities as time elapses.
%The transition probabilities change in different time period $n$.
%Note that the user's movement pattern is unknown to the agent.
\item \emph{Actions}: The action of the agent at each time slot $t$ in period $n$ is a vector $\bm{a}_t^n=[{\cal{L}}_t^n,{\bm{u}}_{1,t}^n, {\bm{u}}_{2,t}^n,$ $\cdots, {\bm{u}}_{U,t}^n]$ that jointly considers the VAP selection and the user association.
The action space $\cal{A}$ is the set of all optional actions.
\item \emph{States}: The state at time slot $t$ in time period $n$ is defined as $\bm{s}_t^n=[\bm{v}_t^n, \bm{w}_t^n]$ that consists of: 
1) the user position $\bm{v}_t^n=[{\bm v}_{1,t}^n,\cdots,{\bm v}_{U,t}^n]$, where $\bm{v}_{j,t}^n$ depends on $\bm{v}_{j,t\!-\!1}^n$ and the movement pattern $\bm{M}^n$ in time period $n$, which is unknown to the central controller and
2) the service state vector $\bm{w}_t^n=[w_{1,t}^n,\cdots,w_{U,t}^n]$ that implies each user whether has been successfully served until time slot $t$.
%3) the set of newly served users ${\cal{W}}_t^n$ at time slot $t$.
The state space $\cal{S}$ is the set of all possible states.
%At the first time slot of each time period, $\bm{v}_1^n$ is generated randomly and the positioning state, service state, and newly server user are initialized as $\bm{p}_1^n=[0,\cdots,0]$, $\bm{w}_1^n=[0,\cdots,0]$, and $\Omega_1^n=\emptyset$, respectively.
\item \emph{Policy}: The policy is the probability of the agent choosing each action at a given state. 
The MPG algorithm uses a deep neural network parameterized by $\bm{\theta}$ to map the input state to the output action.
Then, the policy can be expressed as ${\bm{\pi}}_{\bm{\theta}}(\bm{s}_{t\!-\!1}^n,\bm{a}_t^n)=P(\bm{a}_t^n|\bm{s}_{t\!-\!1}^n)$. 
Based on the policy ${\bm{\pi}}_{\bm{\theta}}$, an execution process in a time period $n$ can be defined as a trajectory $\bm{\tau}^n=\{\bm{s}_0^n,\bm{a}_1^n,\cdots, \bm{s}_{T\!-\!1}^n,\bm{a}_T^n\}$.
\item \emph{Reward}: The benefit of choosing action $\bm{a}_t^n$ at state $\bm{s}_{t\!-\!1}^n$ is ${|{\cal{W}}_t^n(\bm{a}_t^n)|}$.
Therefore, the reward of a trajectory during a time period $n$ is $F^n({\bm{\tau}^n})=\sum\limits_{t = 1}^T {|{\cal{W}}_t^n(\bm{a}_t^n)|}, \forall \bm{a}_t^n \in {\bm{\tau}^n}$. Note that the reward function is equivalent to the number of successfully served users defined in (\ref{reliability}), that is $F^n({\bm{\tau}^n})=R^n({\cal{L}}_{:T}^n,{\bm{u}}_{j,:T}^n)$.
The objective function of problem (\ref{optimization}) that the agent aims to optimize is the average reward function of all time periods $\bar{F}({\bm{\tau}^n})=\frac{1}{N}\sum\limits_{n=1}^N {F^n({\bm{\tau}^n})}=\frac{1}{N}\sum\limits_{n=1}^N {R^n({\cal{L}}_{:T}^n,{\bm{u}}_{j,:T}^n)}$. 
\item \emph{Tasks} : We use a \emph{task} ${\cal{T}}^n$ to refer to the reliability maximization problem $\mathop {\max }\limits_{{\bm{a}}_t^n} F^n({\bm{\tau}^n})$ in each time period $n$.
A task is thus defined as ${\cal{T}}^n=\{\bm{M}^n,F^n({\bm{\tau}^n})\}$. 
For each task ${\cal{T}}^n$, the trajectory $\bm{\tau}^n$ and the corresponding reward $F^n(\bm{\tau}^n)$ are affected by the users' movement pattern $\bm{M}^n$ that is unknown to the agent.
However, the policy ${\bm{\pi}}_{\bm{\theta}}$ is shared by all tasks.
Therefore, the agent must find the effective policy that can quickly adapt to new users' movement patterns.
\end{itemize}

\vspace{-0.5cm}
\subsection{MPG for Optimization of Reliability}
\vspace{-0.1cm}
\label{sec:3.B}
Next, we introduce the entire procedure of training the proposed MPG algorithm.
Our purpose from training MPG is to find the optimal policy that maximizes the reliability of the THz/VLC-enabled wireless VR network over different time periods.
The MPG algorithm enables the trained policy to quickly adapt to the time-varying users' movement patterns.
The intuition behind the proposed MPG is that some of its parameters are task-sensitive while other parameters are broadly applicable to all tasks.
Therefore, the training process of MPG has two steps: 1) task learning step and 2) meta-learning step.
The task learning step enables the MPG to execute the policy gradient on task-sensitive parameters so as to make rapid progress on each new task.
The meta-learning step aims to find the broadly applicable parameters that can improve the performance of all tasks.
The proposed MPG model is trained offline, which means that the MPG model is trained by the trajectories and the corresponding rewards sampled in historical tasks.
Using the historical trajectories and rewards, the MPG model can learn the distribution of the tasks and thus quickly adapt to a new task.
In particular, the trained fast-adaptive MPG model only requires a few iterations of the task learning step to learn the new users' movement pattern so as to solve the new task.
Hence, the proposed algorithm can maximize the reliability of the studied VR network in each specific new time period.
%To applied in the new tasks, the trained fast-adaptive MPG model requires a few iterations of the task learning step for further training so as to maximize the reliability of the studied VR network in each specific new time period.
Specifically, the task learning step and meta-learning step can be given as follows:
\begin{enumerate}
\item \emph{Task learning step}: 
For each task ${\cal{T}}^n$, the agent first collects $K$ trajectories based on a given policy ${\bm{\pi}}_{\bm{\theta}}$.
The set of collected trajectories of task ${\cal{T}}^n$ is ${\cal{D}}^n=\{\bm{\tau}_1^n,\cdots,\bm{\tau}_k^n,\cdots,\bm{\tau}_K^n\}$, where $\bm{\tau}_k^n$ is the trajectory $k$ of task ${\cal{T}}^n$.
To evaluate the policy ${\bm{\pi}}_{\bm{\theta}}$ for maximizing the reliability of the VR network, we define the expected reward of the trajectories in ${\cal{D}}^n$ as
\begin{equation}
\setlength{\abovedisplayskip}{2pt}
\setlength{\belowdisplayskip}{2pt}
\label{expected_reward}
\bar{J}^n(\bm{\theta})=\sum\limits_{\bm{\tau}^n \in {\cal{D}}^n} {P_{\bm{\theta}}({\bm{\tau}^n}){F^n(\bm{\tau}^n)}},
\end{equation}
where $P_{\bm{\theta}}({\bm{\tau}^n})=P({\bm{s}_0^n})\prod\limits_{t = 1}^T {{\bm{\pi}}_{\bm{\theta}}(\bm{s}_{t\!-\!1}^n,\bm{a}_t^n)}P(\bm{s}_{t}^n|\bm{s}_{t\!-\!1}^n,\bm{a}_t^n)$. $P(\bm{s}_{t}^n|\bm{s}_{t\!-\!1}^n,\bm{a}_t^n)$ is the probability of state $\bm{s}_{t\!-\!1}^n$ transitioning to state $\bm{s}_{t}^n$ after taking action $\bm{a}_t^n$, which depends on the movement pattern $\bm{M}^n$.
The goal of optimizing the policy ${\bm{\pi}}_{\bm{\theta}}$ for each task ${\cal{T}}^n$ is to maximize the number of successfully served users in time period $n$, that is 
\begin{equation}
\setlength{\abovedisplayskip}{2pt}
\setlength{\belowdisplayskip}{2pt}
\label{loss_task}
\mathop {\max }\limits_{\bm{\theta}} \bar{J}^n(\bm{\theta}).
\end{equation}
For each task ${\cal{T}}^n$, the policy ${\bm{\pi}}_{\bm{\theta}}$ is updated using the standard gradient ascent method
\begin{equation}
\setlength{\abovedisplayskip}{2pt}
\setlength{\belowdisplayskip}{2pt}
\label{task_policy}
\tilde{\bm{\theta}}^n={\bm{\theta}}+\alpha \nabla_{\bm{\theta}} \bar{J}^n(\bm{\theta}),
\end{equation}
where $\alpha$ is the learning rate that is equal for all tasks and the policy gradient is 
\begin{equation}
\setlength{\abovedisplayskip}{3pt}
\setlength{\belowdisplayskip}{2pt}
\label{task_gradient}
\begin{aligned}
\nabla_{\bm{\theta}} \bar{J}^n(\bm{\theta})=&\sum\limits_{k=1}^K {P_{\bm{\theta}}({\bm{\tau}_k^n}){F^n(\bm{\theta})}} \nabla \log P_{\bm{\theta}}({\bm{\tau}_k^n}),\\
\approx&\frac{1}{K}\sum\limits_{k=1}^K {F^n(\bm{\theta})}\nabla \log P_{\bm{\theta}}({\bm{\tau}_k^n}),\\
=&\frac{1}{K}\sum\limits_{k=1}^K \sum\limits_{t=1}^T {F^n(\bm{\theta})}\nabla \log {\bm{\pi}}_{\bm{\theta}}(\bm{s}_{t\!-\!1}^n,\bm{a}_t^n).
\end{aligned}
\end{equation}
Finally, the agent collects $K'$ trajectories for each task ${\cal{T}}^n$ using the corresponding updated policy ${\bm{\pi}}_{\tilde{\bm{\theta}}^n}$.
Each trajectory set ${\cal{D}}'^{n}=\{\bm{\tau}{'}_1^n,\cdots,\bm{\tau}{'}_{K'}^n\}$ is used to optimize the broadly applicable parameters in the next meta-learning step so as to increase the average number of successfully served user for all tasks.

\item \emph{Meta-learning step}: 
The agent first computes the expected rewards $\bar{J}^n({\tilde{\bm{\theta}}^n})$ of each trajectory set ${\cal{D}}'^{n}$ based on the each updated policy $\tilde{\bm{\theta}}^n$.
To solve the reliability maximization problem (\ref{optimization}), we only need to solve the following optimization problem 
\vspace{-0.1cm}
\begin{equation}
\vspace{-0.1cm}
\label{loss_mata1}
\mathop {\max }\limits_{\bm{\theta}} \frac{1}{N}\sum\limits_{n=1}^N \bar{J}^n({\tilde{\bm{\theta}}^n}).
\end{equation}
Substituting (\ref{task_policy}) into (\ref{loss_mata1}), we have
\vspace{-0.2cm}
\begin{equation}
\vspace{-0.1cm}
\label{loss_mata2}
\mathop {\max }\limits_{\bm{\theta}} \frac{1}{N}\sum\limits_{n=1}^N \bar{J}^n({\bm{\theta}}+\alpha \nabla_{\bm{\theta}} \bar{J}^n(\bm{\theta})).
\end{equation}
Then, to improve the average number of successfully served users for all tasks, the policy ${\bm{\pi}}_{\bm{\theta}}$ is updated by
\vspace{-0.3cm}
\begin{equation}
\vspace{-0.1cm}
\label{meta_gradient}
{\bm{\theta}} \leftarrow {\bm{\theta}}+{\frac{\beta}{N}} \nabla_{\bm{\theta}}\sum\limits_{n=1}^N \bar{J}^n({\bm{\theta}}+\alpha \nabla_{\bm{\theta}} \bar{J}^n(\bm{\theta})),
\end{equation}
where $\beta$ is the learning rate for meta-learning.
Here, note that the meta-learning step is performed over the parameters ${\bm{\theta}}$ instead of the parameters ${\tilde{\bm{\theta}}^n}$ updated in the previous task learning step.
\end{enumerate}
By iteratively running the task learning and the meta-learning step, a locally optimal policy for determining the VAP selection and user association under different users' movement patterns can be obtained.
The specific training process of the proposed MPG algorithm is summarized in \textbf{Algorithm~\ref{algorithm_1}}.

\begin{algorithm}[t]
%\small
\footnotesize
\caption{MPG algorithm for VAP selection and user association.}
\begin{algorithmic}[1]
\STATE \textbf{Input:} The set of VAPs $\cal{V}$, the set of SBSs $\cal{B}$, the user positions $\bm{v}_0^n$, and the transition matrix $\bm{M}^n$.
\STATE \textbf{Initialize:} $\bm{\theta}$ is initially generated randomly, $\bm{w}_0^n=[0,\cdots,0]$, task learning rate $\alpha$, meta-learning rate $\beta$, and the number of iterations $E$.
\FOR {$i = 1 \to E$}
\FORALL {task ${\cal{T}}^n$}
\STATE Collect $K$ trajectories ${\cal{D}}^n=\{\bm{\tau}_1^n,\cdots,\bm{\tau}_k^n,\cdots,\bm{\tau}_K^n\}$ using ${\bm{\pi}}_{\bm{\theta}}$.
\STATE Compute $\nabla_{\bm{\theta}} \bar{J}^n(\bm{\theta})$ using ${\cal{D}}^n$ based on (\ref{task_gradient}).
\STATE Compute parameters $\tilde{\bm{\theta}}^n$ of the adapted policy based on (\ref{task_policy}).
\STATE Collect $K'$ trajectories ${\cal{D}}'^{n}=\{\bm{\tau}{'}_1^n,\cdots,\bm{\tau}{'}_{K'}^n\}$ using ${\bm{\pi}}_{\tilde{\bm{\theta}}^n}$.
\ENDFOR
\STATE Compute $\bar{J}^n({\tilde{\bm{\theta}}^n})$ using each ${\cal{D}}'^{n}$.
\STATE Update the parameters of the policy based on (\ref{meta_gradient}).
\ENDFOR
\end{algorithmic}
\label{algorithm_1}
\end{algorithm}

The optimization problem (\ref{optimization}) is solved once the locally optimal policy of the proposed MPG model that used to determine the VAP selection and user association is obtained.
Since the meta-learning step tends to optimize the broadly applicable parameters for all tasks, the proposed MPG algorithm enables the trained policy to quickly adapt to new tasks.
Once the fast-adaptive VAP selection and user association policy is trained, the central controller can quickly find the locally optimal policy in new time periods.
In particular, for a new task with new users’ movement pattern, using the trained policy as initialization, the central controller can further optimize the policy for the new task by only executing a few iterations.
Hence, the proposed meta learning based algorithms can significantly reduce the training overhead in actual VR scenarios.

\vspace{-0.4cm}
\subsection{Complexity and Overhead of MPG}
\vspace{-0.2cm}
\label{sec:3.C}
Next, we analyze the computational complexity of the proposed MPG algorithm for VAP selection and user association optimization.
The complexity of the MPG algorithm depends on the number of the policy parameter $\bm{\theta}$, which depends on the size of action space $\cal{A}$ and the size of state space $\cal{S}$ \cite{Chen_JSAC_21}.
The action space $\cal{A}$ is a set of all possible VAP selections and user associations.
The number of optional combinations of three VAPs from $V$ VAPs is $C_V^3=\frac{{V!}}{{3!(V - 3)!}}$.
The number of possible user-SBS association depends on the number of users $U$ and the number of SBSs $B$, which is $A_{\max(U,B)}^{\min(U,B)}=\frac{{\max(U,B)!}}{{\left(\max(U,B)-\min(U,B)\right)!}}$.
Hence, the size of $\cal{A}$ will be $\frac{{V!\max(U,B)!}}{{3!(V - 3)!\left(\max(U,B)-\min(U,B)\right)!}}$.
The state space $\cal{S}$ consists of continuous user locations $\bm{v}_t^n$ as well as discrete service state $\bm{w}_t^n$ and newly served users ${\cal{W}}_t^n$.
To ensure the finite state space, we discretize the continuous user positions $\bm{v}_t^n$.
In particular, the considered indoor space $\cal{G}$ is divided into $G$ small grids and the position of the user in each grid is represented by the center of the grid.
The size of service state space $\bm{w}_t^n$ is $U$.
Then, the size of state space $\cal{S}$ is $GU$.
Therefore, the computational complexity of the proposed MPG algorithm can be given as
\begin{equation}
\setlength{\abovedisplayskip}{2pt}
\setlength{\belowdisplayskip}{2pt}
\begin{aligned}
\label{complexity1}
&{\cal{O}}\left\{GU\frac{{V!\max(U,B)!}}{{3!(V\!-\!3)!\left(\max(U,B)\!-\!\min(U,B)\right)!}}\prod\limits_{l = 2}^{L - 1} {{H_l}}\right\}\\
=&{\cal{O}}\left\{UV^3\frac{\max(U,B)!}{\left(\max(U,B)\!-\!\min(U,B)\right)!}\prod\limits_{l = 2}^{L - 1} {{H_l}}\right\},
\end{aligned}
\end{equation}
where $H_l$ is the number of the neurons in layer $l$ of the deep neural network used to train the policy.
From (\ref{complexity1}) we can see that, due to the combinatorial user associations (i.e., $\frac{\max(U,B)!}{\left(\max(U,B)-\min(U,B)\right)!}$), the complexity of the MPG algorithm becomes unacceptably large as the number of users $U$ increases.
%Hence, although the proposed MPG algorithm ensures the local optimal VAP selection and user associations by analyzing the user movement patterns, it can only handle small-scale scenarios with limited VR users.
To this end, we proposed a dual method based MPG (D-MPG) solution in which an action only determines VAP selection. 
Given the VAP selection, the user association can be determined by dual method thus reducing the size of action space of the original MPG algorithm. 
Here, we need to note that the MPG and D-MPG algorithms have their own advantages and drawbacks. 
MPG can converge to a local optimal solution but D-MPG cannot. 
However, D-MPG has a faster convergence compared to the MPG. 
Therefore, one must select the solutions (MPG or D-MPG) based on the implementation requirements such as training time or performance. 
Next, we introduce the D-MPG algorithm.

\vspace{-0.1cm}
\section{Dual Method Based Meta-Learning}
\vspace{-0.1cm}
\label{sec:4}
The components of the D-MPG algorithm are defined as follows: 
\begin{itemize}
\item \emph{Agent}: The agent of the D-MPG is also the central controller.
\item \emph{Actions}: The action of the agent at each time slot $t$ in period $n$ consider the subset of VAPs to select, which is ${\bm{a'}}_t^n={\cal{L}}_t^n$.
The action space is ${\cal{A'}}$.
\item \emph{States}: The state at time slot $t$ in time period $n$ is $\bm{s}_t^n=[\bm{v}_t^n, \bm{w}_t^n]$ and the state space is $\cal{S}$. 
Given the VAP selection ${\bm{a'}}_t^n$, the service state vector $\bm{w}_t^n$ depends on the user association.
The determination of user association using dual method will be specified in Section \ref{sec:4.1}.
\item \emph{Policy}: The policy ${\bm{\pi}}_{\bm{\theta'}}(\bm{s}_{t\!-\!1}^n,\bm{a}_t^n)$ is used to build the relationship between the input state and output action, where ${\bm{\theta'}}$ is the parameter of the deep neural network used to learn the policy.
The trajectory during a time period $n$ based on the policy ${\bm{\pi}}_{\bm{\theta'}}$ can be given as $\bm{\tau}^n=\{\bm{s}_0^n,{\bm{a'}}_1^n,\cdots, \bm{s}_{T\!-\!1}^n,{\bm{a'}}_T^n\}$.
\item \emph{Reward}: The reward of a trajectory in time period $n$ is $F^n({\bm{\tau}^n})=\sum\limits_{t = 1}^T {\left|{\cal{W}}_t^n({\bm{a'}}_t^n,{\bm{u}}_{j,t}^{*n})\right|}=R^n({\cal{L}}_{:T}^n,{\bm{u}}_{j,:T}^{*n}), \forall {\bm{a'}}_t^n \in {\bm{\tau}^n}$, where ${\bm{u}}_{j,t}^{*n}$ is the optimal user association based on the chosen action ${\bm{a'}}_t^n$.
The average reward function of all time periods that the agent aims to optimize is $\bar{F}({\bm{\tau}^n})=\frac{1}{N}\sum\limits_{n=1}^N {F^n({\bm{\tau}^n})}=\frac{1}{N}\sum\limits_{n=1}^N {R^n({\cal{L}}_{:T}^n,{\bm{u}}_{j,:T}^{*n})}$, which is also the objective function of the reliability maximization problem (\ref{optimization}). 
Here, we see that, to maximize the average reward function $\bar{F}({\bm{\tau}^n})$, we need to determine the optimal user association ${\bm{u}}_{j,t}^{*n}$ at each time slot $t$.
\item \emph{Tasks} : Task ${\cal{T}}^n$ is the reliability maximization problem in each time period $n$.
\end{itemize}

From the above definitions, we can see that the only difference between the MPG algorithm and the D-MPG algorithm is \emph{action}. 
In particular, an action of the original MPG jointly determines VAP selection and user association while an action of the D-MPG determines only VAP selection. 
Therefore, the D-MPG can significantly decrease the action space thus improving training complexity and convergence speed.   
Next, we will specify the dual method for user-SBS association optimization.

\vspace{-0.3cm}
\subsection{Optimization of User-SBS Association and Reliability}
\vspace{-0.1cm}
\label{sec:4.1}
Once VAPs are selected, the user-SBS association can be determined based on the user positions to avoid blockages of THz links and meet the transmission delay constraints by solving the optimization problem defined in (\ref{optimization}).
Substituting (\ref{reliability}) into (\ref{optimization}), the user-SBS association and reliability maximization problem with fixed VAP selection ${\cal{L}}_t^n$ can be expressed as
\begin{subequations}
\setlength{\abovedisplayskip}{2pt}
\setlength{\belowdisplayskip}{2pt}
\label{optimization2}
\begin{align}\tag{\theequation}
&{\mathop {\max }\limits_{{\bm{u}}_{j,t}^n} \;\;\sum\limits_{n=1}^N \sum\limits_{t = 1}^T {|{\cal{W}}_t^n({\bm{a'}}_t^n,{\bm{u}}_{j,t}^n)|},}\\
&{\;\;{\rm{s}}.{\rm{t}}.\;\;\;\sum\limits_{i=1}^B u_{ij,t}^n \leqslant 1, \forall j \in {\cal U}_t^n,}\\
&{\;\;\;\;\;\;\;\;\;\;\sum\limits_{j=1}^{U} u_{ij,t}^n \leqslant 1, \forall i \in {\cal B},}\\
&{\;\;\;\;\;\;\;\;\;\;u_{ij,t}^n \in \{0,1\}, \forall i \in {\cal B},\forall j \in {\cal U}_t^n.}
\end{align}
\end{subequations}
From (\ref{user_served}), we can see that users can experience immersive VR services as long as each one of them is successfully served once in each time period.
This means that serving a VR user multiple times in a period cannot improve the reliability of the studied VR network.
Therefore, a problem equivalent to (\ref{optimization2}) is
\begin{subequations}
\label{optimization3}
\begin{align}\tag{\theequation}
&{\mathop {\max }\limits_{{\bm{u}}_{j,t}^n} \;\;\sum\limits_{n=1}^N \sum\limits_{t = 1}^T {|{\cal{W}}_t^n({\bm{a'}}_t^n,{\bm{u}}_{j,t}^n)|},}\\
&{\;\;{\rm{s}}.{\rm{t}}.\;\;\;{\rm{(\ref{optimization2}a)-(\ref{optimization2}c)}},}\\
&{\;\;\;\;\;\;\;\;\;\;\sum\limits_{t=1}^{T} \sum\limits_{i=1}^B u_{ij,t}^n \leqslant 1, \forall j \in {\cal U}_t^n,}
\end{align}
\end{subequations}
where (\ref{optimization3}b) indicates that each VR user can be served at most once in a time period. 
Based on (\ref{optimization3}b), the newly served user at each time slot defined in (\ref{user_served}) can be simplified to
\begin{equation}
\label{user_served2}
\begin{aligned}
{\cal{W}}_t^n\!({\bm{a'}}_t^n\!,\!{\bm{u}}_{j,t}^n)\! &\!=\!\{ j|{w_{j,t}^n}({\bm{a'}}_t^n\!,\!{\bm{u}}_{j,t}^n)\!=\!1\!,{w_{j,t\!-\!1}^n}({\bm{a'}}_{\!t\!-\!1}^n\!,\!{\bm{u}}_{j,t\!-\!1}^n)\!=\!0\}\!,\\
&\!\xlongequal{(\ref{optimization3}b)}\! \{ j|{w_{j,t}^n}({\bm{a'}}_t^n\!,\!{\bm{u}}_{j,t}^n) \!=\! 1\}.
\end{aligned}
\end{equation}
This is because ${w_{j,t}^n}({\bm{a'}}_t^n\!,\!{\bm{u}}_{j,t}^n)\!=\!1$ and ${w_{j,t\!-\!1}^n}({\bm{a'}}_{\!t-1}^n\!,\!{\bm{u}}_{j,t-1}^n)\!=\!0$ must always be satisfied simultaneously with the additional service constraint (\ref{optimization3}b).
Then, substituting (\ref{service_state}) into (\ref{user_served2}), we have
\begin{equation}
\label{num_user_served}
\begin{aligned}
|{\cal{W}}_t^n({\bm{a'}}_t^n\!,\!{\bm{u}}_{j,t}^n)| &\!= \!\left|\{ j|{w_{j,t}^n}({\bm{a'}}_t^n\!,\!{\bm{u}}_{j,t}^n) \!=\! 1\}\right|,\\
&\!=\!\!\sum\limits_{j=1}^U\!\left(\!({p_{j,t}^n}(\!{\bm{a'}}_t^n\!){h_{j,t}^n}(\!\bm{u}_{j,t}^n\!))\!\vee\! {w_{j,t-1}^n}(\!{\bm{a'}}_{\!t\!-\!1}^n,{\bm{u}}_{j,t\!-\!1}^n\!)\!\right)\!,\\
&\!\xlongequal{(\ref{optimization3}b)}\! \sum\limits_{j=1}^U {p_{j,t}^n}({\bm{a'}}_t^n){h_{j,t}^n}(\bm{u}_{j,t}^n).
\end{aligned}
\end{equation}
Here, due to (\ref{optimization3}b), the service state ${w_{j,t-1}^n}({\cal{L}}_{t-1}^n\!,\!{\bm{u}}_{j,t-1}^n)\!=\!0$ at time slot $t-1$ must be satisfied if we have $({p_{j,t}^n}(\!{\bm{a'}}_t^n\!){h_{j,t}^n}(\!\bm{u}_{j,t}^n\!))\!=\!1$ at time slot $t$. 
Hence, (\ref{num_user_served}) that represents the number of served user at each time slot $t$ given the selected VAPs ${\bm{a'}}_t^n$ is obtained.

Since optimizing the user association in each time period $n$ is independent, problem (\ref{optimization3}) can be decoupled into multiple subproblems.
In addition, due to the binary variable ${\bm{u}}_{j,t}^n$, the optimization problem in (\ref{optimization3}) is hard to solve. 
Hence, we temporarily adopt the fractional user association relaxation, where association variable ${\bm{u}}_{j,t}^n$ can take on any real value in $[0,1]$. 
As proved in \cite{UAV_VLC}, although the feasible region of ${\bm{u}}_{j,t}^n$ is relaxed to be continuous, the optimal solution of the relaxed problem also meets the integer constraint.
Therefore, the relaxation does not cause any loss of optimality to the final solution of problem (\ref{optimization3}).
Then, for each time period $n$, the reliability maximization subproblem can be formulated as follows:

\begin{subequations}
\setlength{\abovedisplayskip}{3pt}
\setlength{\belowdisplayskip}{2pt}
\label{optimization4}
\begin{align}\tag{\theequation}
&{\mathop {\max }\limits_{{\bm{u}}_{j,t}^n} \;\;\sum\limits_{t=1}^T\sum\limits_{j=1}^U {p_{j,t}^n}({\bm{a'}}_t^n){h_{j,t}^n}(\bm{u}_{j,t}^n),}\\
&{\;\;{\rm{s}}.{\rm{t}}.\;\;\sum\limits_{i=1}^B u_{ij,t}^n \leqslant 1, \forall j \in {\cal U'}_t^n,}\\
&{\;\;\;\;\;\;\;\;\;\sum\limits_{j=1}^{U'} u_{ij,t}^n \leqslant 1, \forall i \in {\cal B},}\\
&{\;\;\;\;\;\;\;\;\;\sum\limits_{t=1}^{T} \sum\limits_{i=1}^B u_{ij,t}^n \leqslant 1, \forall j \in {\cal U'}_t^n,}\\
&{\;\;\;\;\;\;\;\;\;u_{ij,t}^n \in [0,1], \forall i \in {\cal B},\forall j \in {\cal U'}_t^n,}\\
&{\;\;\;\;\;\;\;\;\;u_{ij,t}^n =0, \forall i \in {\cal B},\forall j \in {\cal U}_t^n\backslash {\cal U'}_t^n,}
\end{align}
\end{subequations}
where ${\cal U'}_t^n=\{j|p_{j,t}^n({\bm{a'}}_t^n)=1\}$ represents the set of users that are successfully localized by the set of VAPs ${\cal{L}}_t^n$ at time slot $t$ in time period $n$ and $U'$ is the number of users in ${\cal U'}_t^n$.
Note that problem (\ref{optimization4}) becomes convex after the binary variable ${\bm{u}}_{j,t}^n$ is relaxed.
Here, we ignore the blockages of THz links caused by the users that are not successfully localized by the set of VAPs ${\cal{L}}_t^n$ (i.e., $j \in {\cal U}_t^n\backslash {\cal U'}_t^n$).

Due to constraint (\ref{optimization4}c), all time slots are coupled in problem \eqref{optimization4}.
To simplify problem \eqref{optimization4}, we use the dual method to decouple problem \eqref{optimization4} into multiple subproblems.
The dual problem of \eqref{optimization4} can be given by
\begin{subequations}
\setlength{\abovedisplayskip}{3pt}
\setlength{\belowdisplayskip}{2pt}
\label{optimization5}
\begin{align}\tag{\theequation}
&{\mathop {\max }\limits_{{\bm{u}}_{j,t}^n} \;\;\sum_{t=1}^T\sum\limits_{j=1}^U {p_{j,t}^n}({\bm{a'}}_t^n){h_{j,t}^n}(\bm{u}_{j,t}^n)}
\!+\!\sum_{j=1}^U\lambda_j\left(1-\sum\limits_{t=1}^{T} \sum\limits_{i=1}^B u_{ij,t}^n\right)\\
&{\;\;{\rm{s}}.{\rm{t}}.\;\;\;\sum\limits_{i=1}^B u_{ij,t}^n \leqslant 1, \forall j \in {\cal U'}_t^n,}\\
&{\;\;\;\;\;\;\;\;\;\;\sum\limits_{j \in {\cal U'}_t^n} u_{ij,t}^n \leqslant 1, \forall i \in {\cal B},}\\
&{\;\;\;\;\;\;\;\;\;\;u_{ij,t}^n \in \{0,1\}, \forall i \in {\cal B},\forall j \in {\cal U'}_t^n,}\\
&{\;\;\;\;\;\;\;\;\;\;u_{ij,t}^n =0, \forall i \in {\cal B},\forall j \notin {\cal U'}_t^n,}
\end{align}
\end{subequations}
where $\lambda_j$ is the dual variable associated with constraint (\ref{optimization4}c).
According to \cite{bertsekas2009convex}, there is no gap between convex problem (\ref{optimization4}) and its dual problem (\ref{optimization5}).
Hence, the solution of problem (\ref{optimization5}) is the same as the solution of problem (\ref{optimization4}) and is also the solution of the original problem (\ref{optimization3}).

Since both the objective function and constraints in \eqref{optimization5} can be decoupled, the reliability maximization subproblem at time slot $t$ in period $n$ can be given by
\vspace{-0.1cm}
\begin{subequations}
\label{optimization6}
\begin{align}\tag{\theequation}
&{\mathop {\max }\limits_{{\bm{u}}_{j,t}^n} \;\;\sum\limits_{j=1}^U \left({p_{j,t}^n}({\bm{a'}}_t^n){h_{j,t}^n}(\bm{u}_{j,t}^n)-\lambda_j\sum\limits_{i=1}^B u_{ij,t}^n\right),}\\
&{\;\;{\rm{s}}.{\rm{t}}.\;\;\;\sum\limits_{i=1}^B u_{ij,t}^n \leqslant 1, \forall j \in {\cal U'}_t^n,}\\
&{\;\;\;\;\;\;\;\;\;\;\sum\limits_{j \in {\cal U'}_t^n} u_{ij,t}^n \leqslant 1, \forall i \in {\cal B},}\\
&{\;\;\;\;\;\;\;\;\;\;u_{ij,t}^n \in \{0,1\}, \forall i \in {\cal B},\forall j \in {\cal U'}_t^n,}\\
&{\;\;\;\;\;\;\;\;\;\;u_{ij,t}^n =0, \forall i \in {\cal B},\forall j \notin {\cal U'}_t^n,}
\end{align}
\end{subequations}
which can be solved by the Hungarian algorithm \cite{chen_FL}.
The dual variable $\lambda_j$ can be updated by using the gradient method \cite{bertsekas2009convex}.
The update procedure is given by
\vspace{-0.1cm}
\begin{eqnarray}
\label{co2eq5}
\lambda_j =\left[\lambda_j - \phi\left(1-\sum\limits_{t=1}^{T} \sum\limits_{i=1}^B u_{ij,t}^n\right)\right], \quad \forall j \in\mathcal U,
\end{eqnarray}
where $\phi>0$ is a dynamically chosen step-size sequence and $[x]^+=\max\{x,0\}$.
The specific process of using the proposed D-MPG algorithm to determine the VAP selection and user association is summarized in \textbf{Algorithm~\ref{algorithm_2}}.

\begin{algorithm}[t]
%\small
\footnotesize
\caption{Dual method based algorithm for VAP selection and user association.}
\begin{algorithmic}[1]
\STATE \textbf{Input:} The set of VAPs $\cal{V}$, the set of SBSs $\cal{B}$, the user positions $\bm{v}_0^n$, and the transition matrix $\bm{M}^n$.
\STATE \textbf{Initialize:} ${\bm{\theta'}}$ is initially generated randomly, $\bm{w}_0^n=[0,\cdots,0]$, task learning rate $\alpha$, meta-learning rate $\beta$, the number of iterations $E$, the dual variable $\lambda_j$, and step size $\phi$.
\FOR {$i = 1 \to E$}
\FORALL {task ${\cal{T}}^n$}
\FORALL {time slot $t$}
\STATE Choose ${\bm{a'}}_t^n$ using ${\bm{\pi}}_{\bm{\theta'}}$.
\STATE Determine ${\bm{u}}_{j,t}^n$ based on the solution of (\ref{optimization2}) using the dual method.
\ENDFOR
\STATE Collect $K$ trajectories ${\cal{D}}^n=\{\bm{\tau}_1^n,\cdots,\bm{\tau}_k^n,\cdots,\bm{\tau}_K^n\}$.
\ENDFOR
\STATE Update the parameters ${\bm{\theta'}}$ of the policy as done in \textbf{Algorithm~\ref{algorithm_1}}.
\ENDFOR
\end{algorithmic}
\label{algorithm_2}
\end{algorithm}

Once the user-SBS association is determined based on the user positions measured by the selected VAPs, the \emph{states} and \emph{reward} of the D-MPG algorithm can be obtained.
Then, the D-MPG algorithm can train the VAP selection policy offline using the trajectories and the corresponding rewards sampled in historical tasks according to the training process defined in Section \ref{sec:3.B}.
Here, we need to point out that the users that are not successfully localized may block the THz links that are established according to the user association obtained by the D-MPG algorithm thus damaging the reliability of the THz/VLC-enabled VR network.
This is because, at each time slot, the user association optimization problem (\ref{optimization2}) can only consider the users who is successfully localized using the selected VAPs ${\bm{a'}}_t^n$.
However, compared to the original MPG algorithm, the proposed D-MPG algorithm can control the tradeoff between algorithm processing time or space and reliability gain achieved by the algorithm.
%Note that, at each time slot $t$, only the users who satisfy $p_{j,t}^n({\cal{L}}_t^n)=1$ can be successfully located using the selected VAPs and the location-unknown users may block the THz links established according to the optimized user association. 

\vspace{-0.4cm}
\subsection{Complexity and Overhead of the Proposed Algorithms}
\vspace{-0.1cm}
\label{sec:4.2}
The complexity of the D-MPG algorithm lies in training the policy for VAP selection and optimizing the user association by the dual method.
The complexity for training the VAP selection policy depends on the size of action space ${\cal{A'}}$ and the size of state space $\cal{S}$.
Since an action of the proposed D-MPG algorithm only determines VAP selection, the size of ${\cal{A'}}$ is $C_V^3=\frac{{V!}}{{3!(V - 3)!}}$. 
As analyzed in Section \ref{sec:3.C}, the size of $\cal{S}$ is still $GU$.
Given the VAP selection, the complexity of optimizing user association lies in solving the dual problem of user association optimization problem using Hungarian algorithm whose worst-case complexity is $U^2B$.
To determine the user association for $N$ time slot in a time period, the complexity is $NU^2B$.
Therefore, the complexity of the D-MPG algorithm can be given as
\vspace{-0.1cm}
\begin{equation}
\label{complexity2}
{\cal{O}}\left\{ \frac{{GV!U}}{{3!(V - 3)!}}\prod\limits_{l = 2}^{L - 1} {{H_l}}+ NU^2B\right\}={\cal{O}}\left\{ UV^3\prod\limits_{l = 2}^{L - 1} {{H_l}}+NU^2B\right\}.
\end{equation}
Compared with (\ref{complexity1}), the complexity of solving the proposed reliability maximization problem is significantly reduced.

The proposed MPG and D-MPG models are trained offline, which means that the models are trained during the idle time of the central controller using the trajectories and the corresponding rewards sampled in historical tasks.
Hence, we ignored the time and energy consumption of the training process of the proposed algorithms.
In actual VR applications, we first train the proposed model offline.
The trained model is considered as an initialization model for new learning tasks and can be directly implemented without training. 
Meanwhile, when implementing a new task, the proposed meta learning algorithm can further improve its pre-trained initialization model to achieve better reliability of the studied network using a small number of iterations. 

\vspace{-0.2cm}
\section{Simulation Results and Analysis}
\vspace{-0.1cm}
\label{sec:5}
For our simulations, a $6$~m $\times$ $6$~m square room is considered with $D=7$ VAPs and $B=7$ SBSs evenly distributed at a fixed hight of $Z=3$~m.
$U=20$ wireless VR users are initially randomly distributed in the room and move according to users' movement patterns ${\bm{M}}^n$ generated based on a given distribution in each time period $n$. 
For comparison purposes, we consider the trust region policy optimization algorithm (TRPO) as the baseline scheme. 
TRPO is a widely used RL algorithm that has been proven to always converge to the optimal or local optimal solution in a static environment \cite{TRPO}.
All statistical results are averaged over a large number of independent runs.
Other parameters are listed in Table \uppercase\expandafter{\romannumeral1} \cite{3D_THz, THz_VTC}. 
\begin{table}\footnotesize
\newcommand{\tabincell}[2]{\begin{tabular}{@{}#1@{}}#1.6\end{tabular}}
\renewcommand\arraystretch{1}
\caption[table]{{System parameters}}
\centering
\begin{tabular}{|c|c|c|c|c|c|}
\hline
\!\textbf{Parameters}\! \!\!& \textbf{Value} &\! \textbf{Parameters} \!& \textbf{Value}\\
\hline
$\Psi_{\frac{1}{2}}$ & $60^ \circ$ & $P$ & 5 dBm\\
\hline
$f$ &  1.05 THz & $S$ & 20 Mbit \\
\hline
$\iota_{\rm{T}}$ & 0.1 & $\iota_{\rm{R}}$ & 0.1 \\
\hline
${\varphi _{\rm{H}}^{\rm{T}}}$ & $10^ \circ$ & ${\varphi _{\rm{V}}^{\rm{T}}}$ & $10^ \circ$ \\
\hline
 ${\varphi _{\rm{H}}^{\rm{R}}}$ & $33^ \circ$ & ${\varphi _{\rm{V}}^{\rm{R}}}$& $33^ \circ$ \\
\hline
$K_BT_e$ & $-174$ dBm/Hz & $K(f)$& 0.07512 m$^{-1}$ \\
\hline
 $K$ & 50 & $K'$ & 10\\
 \hline
 $\alpha$ & 0.1 & $\beta$ &0.01  \\
 \hline
 \end{tabular}
 \vspace{-0.3cm}
\end{table}

\begin{figure}[t]
\centering
\vspace{-0.3cm}
\setlength{\belowcaptionskip}{-1cm}
\setlength{\abovecaptionskip}{0cm}
\includegraphics[width=9.5cm]{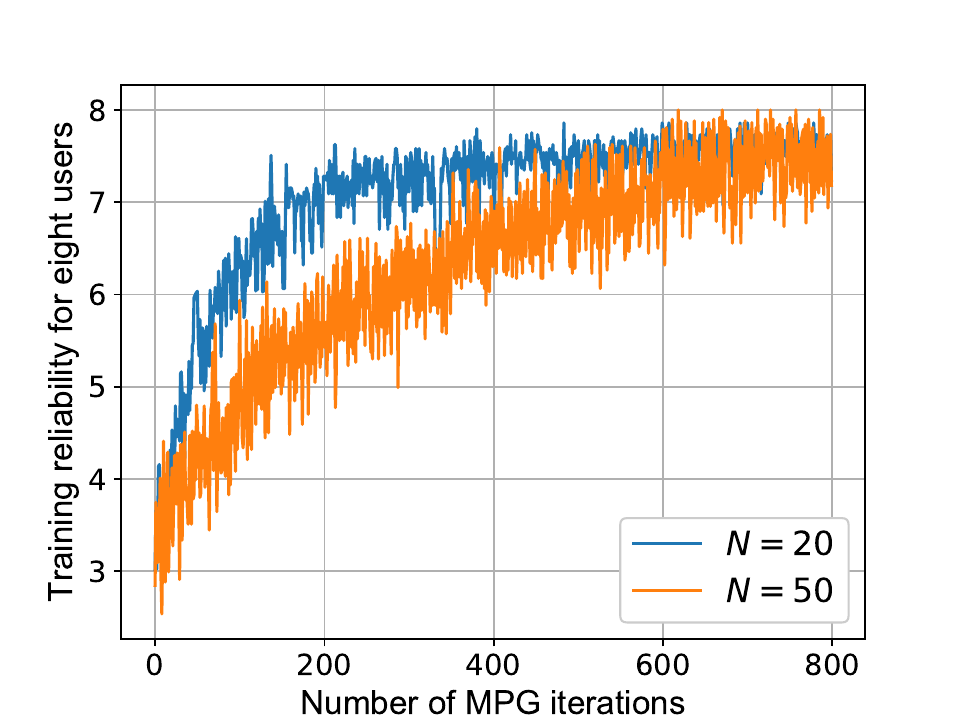}
%\captionsetup{labelfont={blue},textfont={blue}}
\caption{The reliability of the THz/VLC-enabled VR network as the number of iterations of the MPG algorithm varies.}
\label{fig2}
\end{figure}

Fig.~\ref{fig2} shows how the reliability of a THz/VLC-enabled VR network changes as the number of iterations of the MPG algorithm varies.
Here, a VR scenario with 8 users is considered for the proposed MPG algorithm due to the huge computational overhead.
The MPG model is trained for $N=20$ tasks and $N=50$ tasks to obtain a fast-adaptive policy for VAP selection and user association, respectively.
In Fig.~\ref{fig2}, we can see that the proposed MPG algorithm can converge and effectively solve the reliability maximization problem (\ref{optimization}).
This is due to the fact that the proposed MPG algorithm can analyze the distribution of the dynamic users' movement patterns so as to update the policy to increase the average number of the successfully served users for all tasks.
Fig.~\ref{fig2} also shows that the training process of $N=20$ tasks is more stable and converges faster than the training process of $N=50$ tasks.
This is because fewer tasks are more likely to find update gradients that work for most of the tasks in the meta-learning step.

\begin{figure}[t]
\vspace{-1cm}
\centering
\setlength{\belowcaptionskip}{-0.8cm}
\setlength{\abovecaptionskip}{0cm}
\includegraphics[width=9.5cm]{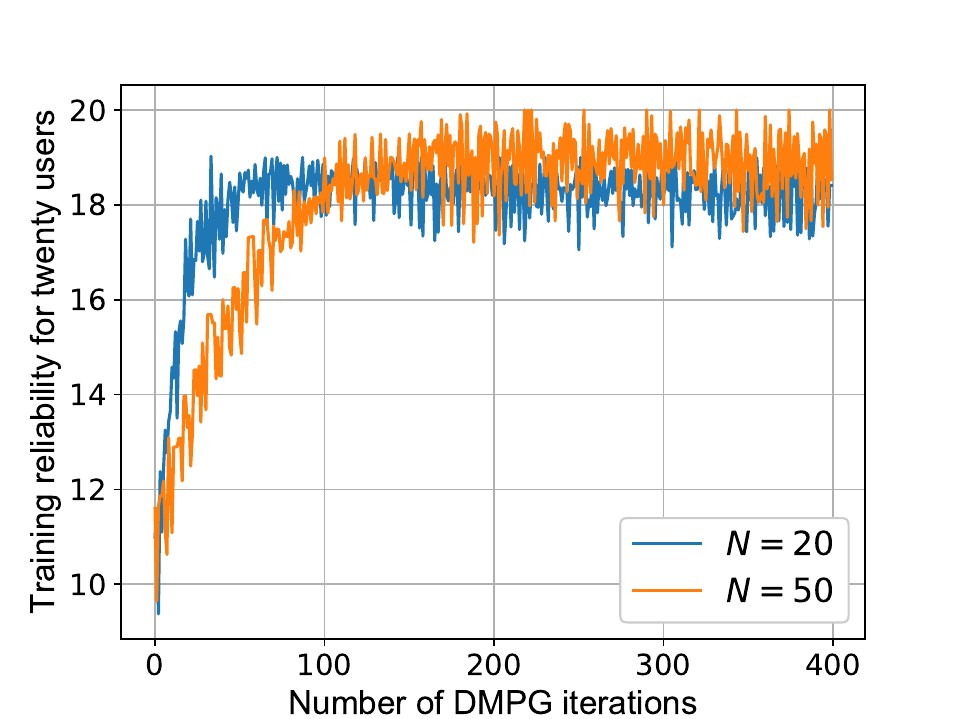}
\caption{The convergence of the D-MPG algorithm in a network with 20 users.}
\label{fig3}
\end{figure}

Fig.~\ref{fig3} shows the convergence of the D-MPG algorithm in a network with 20 users.
%{\color{blue}{In Fig. 3, the proposed D-MPG algorithm trains a fast-adaptive VAP selection policy for a VR network with 20 users.}}
%This is due to the fact that, in the D-MPG algorithm, the user association is determined by the dual method, thus reducing the computational complexity.
In Fig.~\ref{fig3} we can also see that, using the D-MPG algorithm, $N=20$ tasks require approximately 30\% the number of iterations needed to reach convergence compared for the case with $N=50$ tasks.
This is because the D-MPG algorithm needs to analyze the users' movement patterns of all tasks so as to train the locally optimal VAP selection policy.
%Fig.~\ref{fig3} also shows that the D-MPG algorithm for $N=20$ tasks and $N=50$ tasks finally converges at the same reliability which indicates that the number of tasks will not affect the reliability of the THz/VLC-enabled VR network.

\begin{figure}[t]
\centering
\setlength{\belowcaptionskip}{-1cm}
\setlength{\abovecaptionskip}{0cm}
\includegraphics[width=9.5cm]{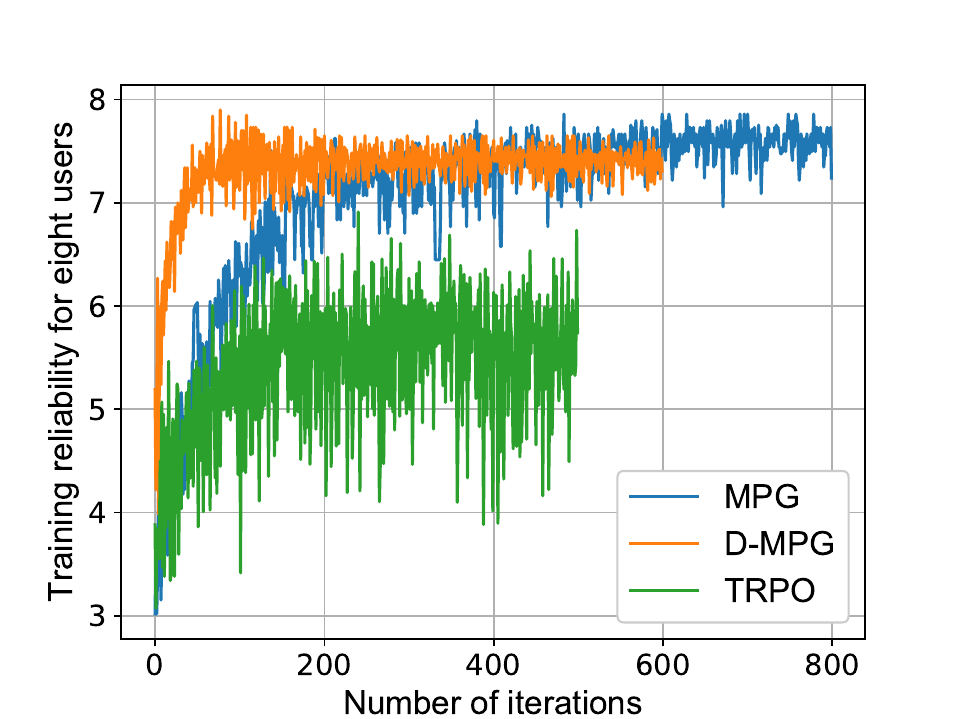}
%\captionsetup{labelfont={blue},textfont={blue}}
\caption{The training process of all considered RL algorithms.}
\label{fig4}
\end{figure}

Fig.~\ref{fig4} shows the training reliability for eight VR users resulting from all of the considered RL algorithms.
In Fig.~\ref{fig4}, we can see that, compared with the baseline TRPO algorithm that cannot converge in presence of dynamic users' movement patterns, the proposed MPG and D-MPG algorithm can reach convergence after approximately 400 and 200 iterations, respectively.
Meanwhile, compared to the TRPO algorithm, the proposed MPG algorithm and D-MPG algorithm can achieve 33.3\% and 29.8\% gains in terms of reliability that is averaged over the last 400 iterations.
This stems from the fact that the proposed MPG and D-MPG algorithms can build a relationship between the dynamic users' movement patterns by meta-learning step.
Fig. 4 also shows that the reliability of the studied VR network achieved by the MPG algorithm is 2.7\% higher than the reliability achieved by the D-MPG algorithm.
This is because the MPG algorithm determines the user-SBS associations based on the learned users' movement patterns and hence, it optimizes user-SBS association while considering all VR user locations.
However, the D-MPG algorithm optimizes the user-SBS associations for the successfully localized users without considering the users that are not successfully localized.
Fig.~\ref{fig4} also shows that, compared with the MPG algorithm, the D-MPG algorithm converges faster.
This is due to the fact that the policy of the D-MPG algorithm only needs to learn how to select VAPs, while the policy of the MPG algorithm needs to learn how to determine the VAP selection and the user association.

\begin{figure}[t]
\vspace{-1cm}
\centering
\setlength{\belowcaptionskip}{-0.8cm}
\setlength{\abovecaptionskip}{0cm}
\includegraphics[width=9.5cm]{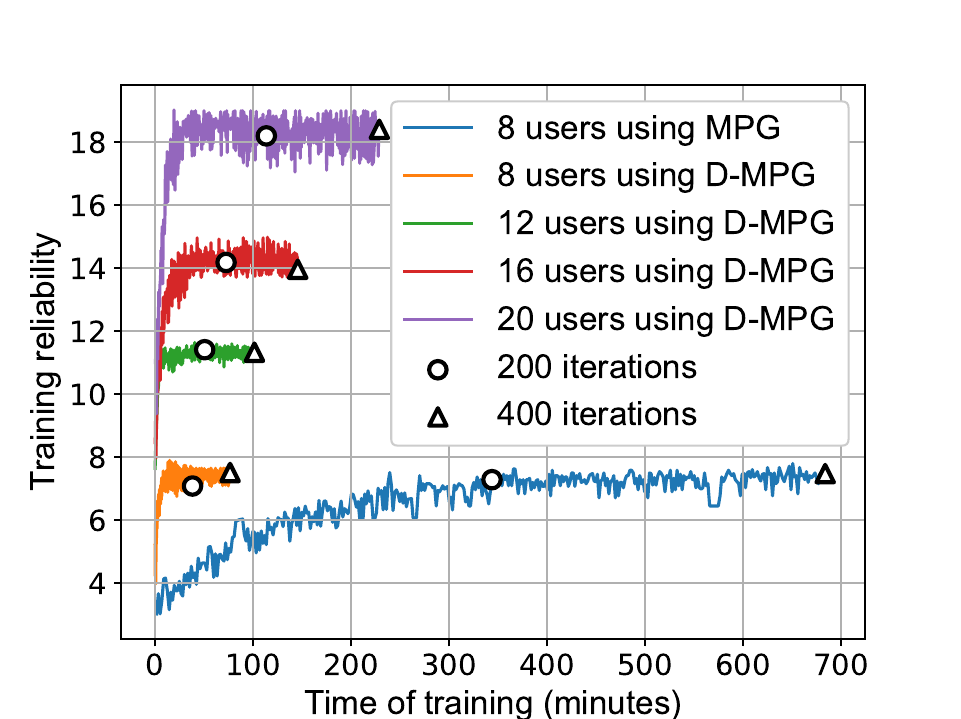}
%\captionsetup{labelfont={blue},textfont={blue}}
\caption{The training time of the proposed MPG algorithm and DMPG algorithm.}
\label{fig5}
\end{figure}

In Fig.~\ref{fig5}, we show how the training time of the proposed MPG algorithm and D-MPG algorithm will change as the number of users varies.
From Fig.~\ref{fig5} we can see that the D-MPG algorithm for 8 users can yield up to 88.7\% reduction in terms of training time compared with the MPG algorithm.
This gain stems from the fact that the use of the dual method to determine the user association can significantly reduce the computational complexity of the D-MPG algorithm.
In Fig.~\ref{fig5}, we can also find that, as the number of VR users increases, the training time of the D-MPG algorithm increases.
This is because the complexity of the D-MPG algorithm is a function of the number of VR users as analyzed in Section \ref{sec:4.2}.

\begin{figure}[t]
\centering
\setlength{\belowcaptionskip}{-0.8cm}
\setlength{\abovecaptionskip}{0cm}
\includegraphics[width=8.7cm]{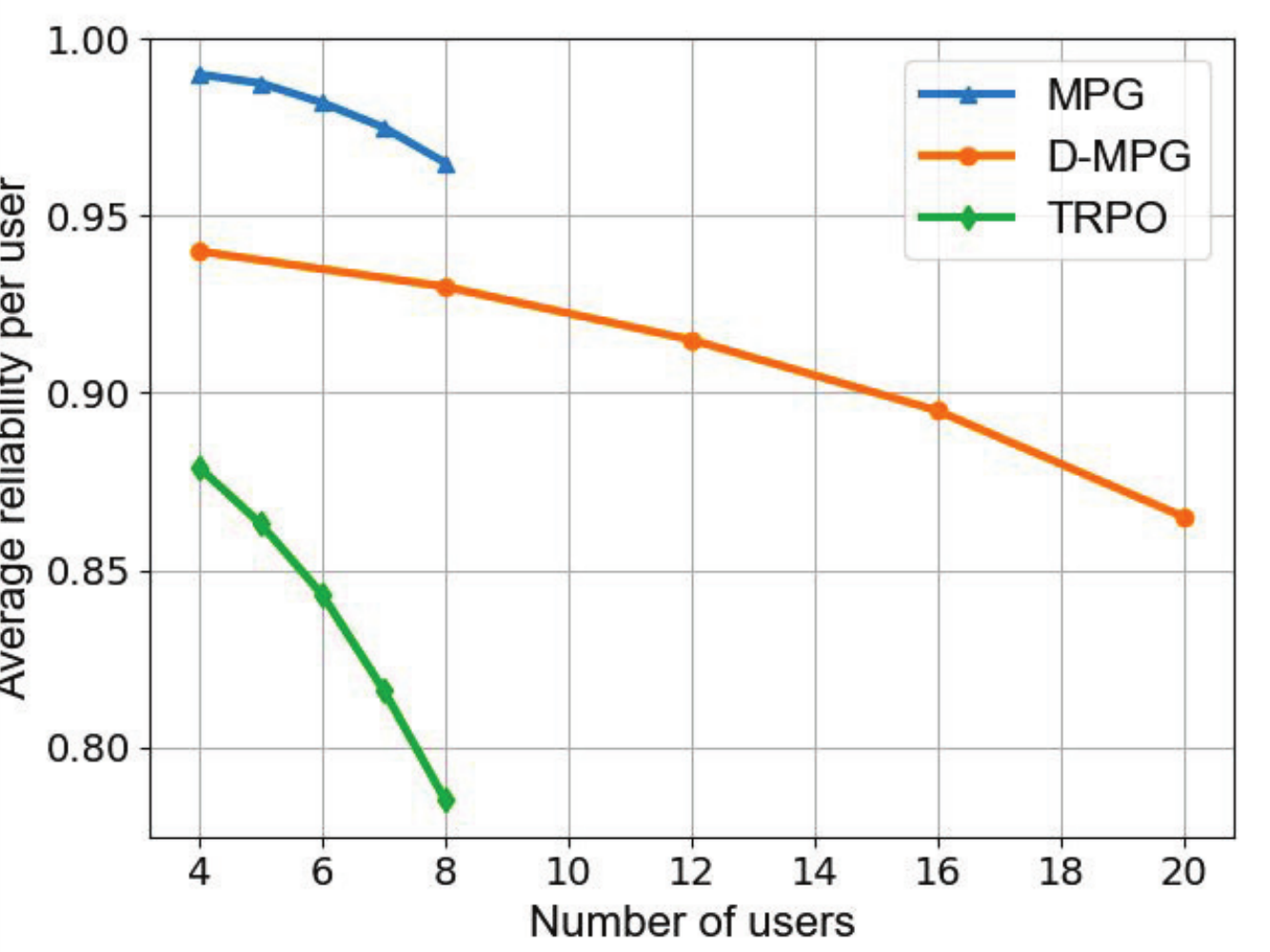}
%\captionsetup{labelfont={blue},textfont={blue}}
\caption{The average reliability per user as the number of users varies.}
\label{fig6}
\end{figure}

Fig.~\ref{fig6} shows how the average reliability per user changes as the number of users varies.
In Fig.~\ref{fig6}, the number of users varies from 4 to 8 for the MPG algorithm and the baseline TRPO algorithm and the number of users varies from 4 to 20 for the D-MPG algorithm.
From Fig.~\ref{fig6} we can see that the average reliability per user of all the considered algorithms decreases as the number of users increases.
This is due to the fact that, as the number of users in a given area increases, the probability of each user blocking the VLC or THz links increases.
In addition, $B$ SBSs can serve a limited number of users in $T$ time slots (at most $BT$ users), hence the average reliability per user should decrease as the number of users increases.
Fig.~\ref{fig6} also shows that the proposed MPG algorithm and D-MPG algorithm can respectively yield up to 17.1\% and 14.5\% improvements in terms of the average reliability per user compared to the TRPO algorithm.
This is because the proposed MPG and D-MPG algorithms can analyze and quickly adapt the dynamic users' movement patterns so as to determine the VAP selection and user association that can maximize the reliability of the studied VR network.
In Fig.~\ref{fig6} we can also observe that, even in a dense environment (36 m$^2$ with 20 users), the proposed D-MPG algorithm can guarantee an average reliability per user of more than 0.9.
This indicates that the studied THz/VLC network can provide reliable wireless VR services using the proposed algorithms.

\begin{figure}[t]
%\vspace{-1cm}
\centering
\setlength{\belowcaptionskip}{-1cm}
\setlength{\abovecaptionskip}{0cm}
\includegraphics[width=8.7cm]{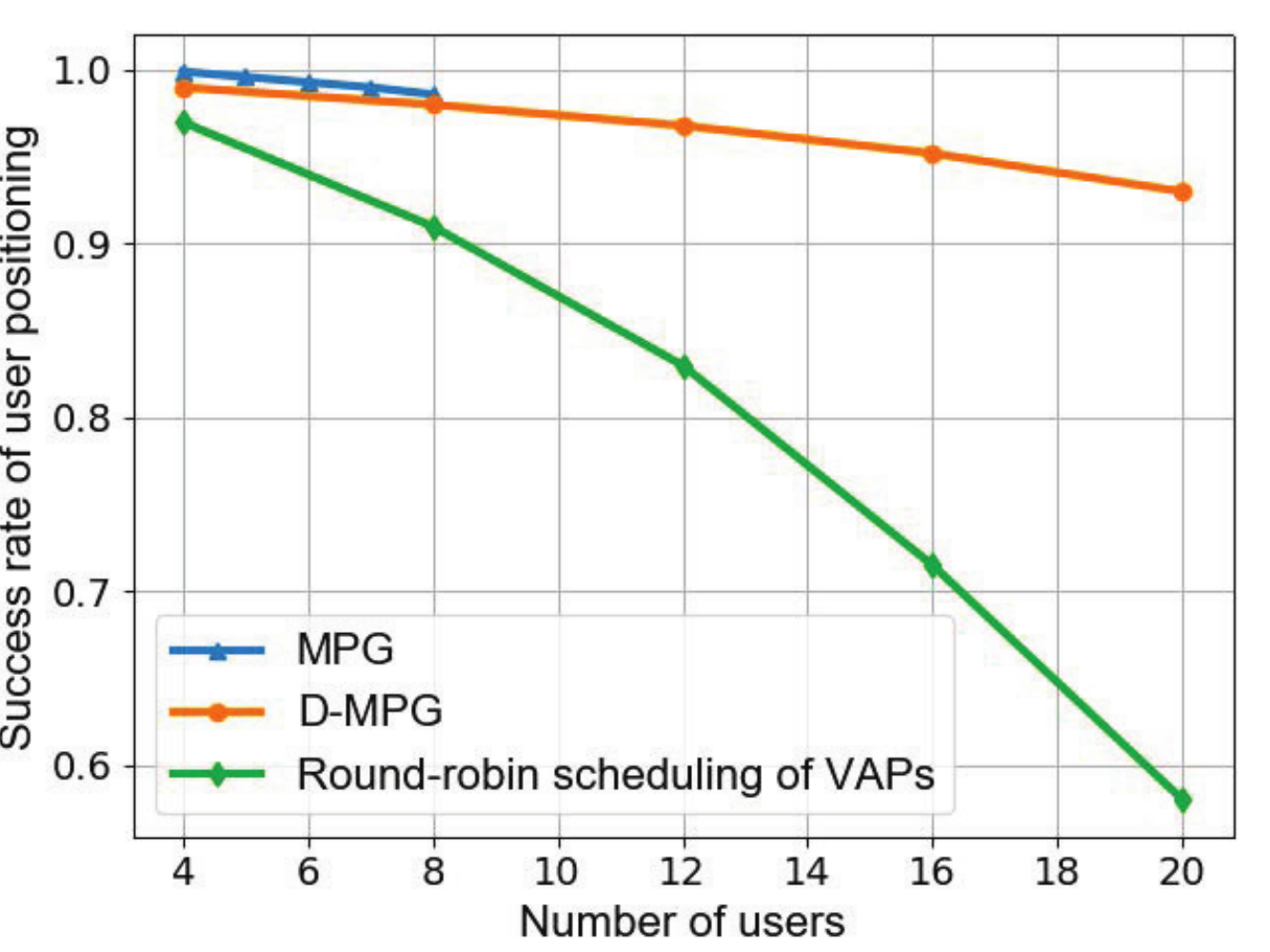}
%\captionsetup{labelfont={blue},textfont={blue}}
\caption{The success rate of the user positioning as the number of users varies.}
\label{fig9}
\end{figure}

Fig. \ref{fig9} shows how the number of users that can be successfully localized changes as the total number of users varies.
In Fig. \ref{fig9}, a round-robin scheduling policy for VAP selection is considered as a baseline.
From Fig. \ref{fig9}, we can see that the success rate of user positioning decreases as the number of users increases.
This is due to the fact that, as the number of users in a given area increases, the probability of the VLC links being blocked by other users increases.
Fig. \ref{fig9} also shows that the proposed MPG algorithm and D-MPG algorithm can respectively yield up to 5.6\% and 23.9\% improvements in terms of the success rate of the user positioning compared to the baseline.
The reason is that the proposed algorithms can analyze the users’ movement patterns and thus optimizing the VAP selection policy to avoid blockages over VLC links.

\begin{figure}[t]
\centering
\vspace{-1cm}
\setlength{\belowcaptionskip}{-1cm}
\setlength{\abovecaptionskip}{0cm}
\includegraphics[width=9cm]{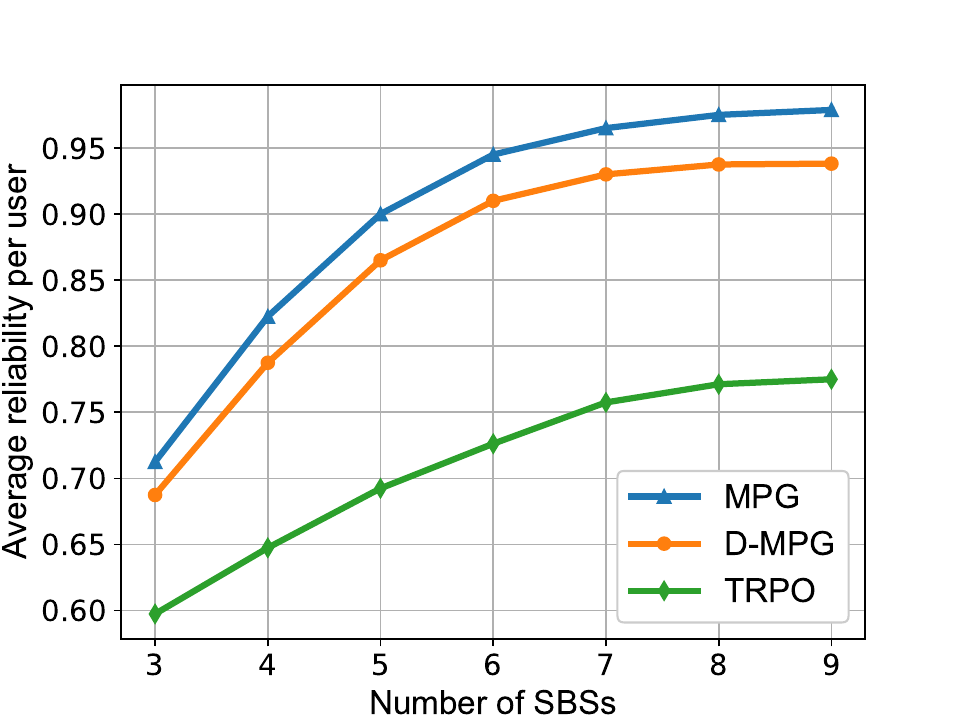}
%\captionsetup{labelfont={blue},textfont={blue}}
\caption{The average reliability per user as the number of SBSs varies ($V=7$).}
\label{fig7}
\end{figure}

Fig.~\ref{fig7} shows how the average reliability per user changes as the number of SBSs varies in a room with 7 VAPs.
From this figure, we observe that, as the number of SBSs increases, the average reliability per user of all algorithms increases since more SBSs can increase the access probability of VR users.
Fig.~\ref{fig7} also shows that the MPG and D-MPG algorithms can achieve up to 26.8\% and 21.9\% gains in terms of average reliability per user compared to the TRPO algorithm, respectively.
Meanwhile, in Fig.~\ref{fig7}, as the number of SBSs increases, the average reliability per user resulting from the proposed algorithms increases more significantly than that of the TRPO algorithm.
This is due to the fact that the proposed algorithms determine the actions based on the analysis of the users' movement patterns, thus enabling the SBSs to cooperatively provide services to VR users.
In consequence, the reliability of the THz/VLC-enabled VR network is significantly improved using the proposed algorithms.

\begin{figure}[t]
\centering
\vspace{-1cm}
\setlength{\belowcaptionskip}{-0.8cm}
\setlength{\abovecaptionskip}{0cm}
\includegraphics[width=9cm]{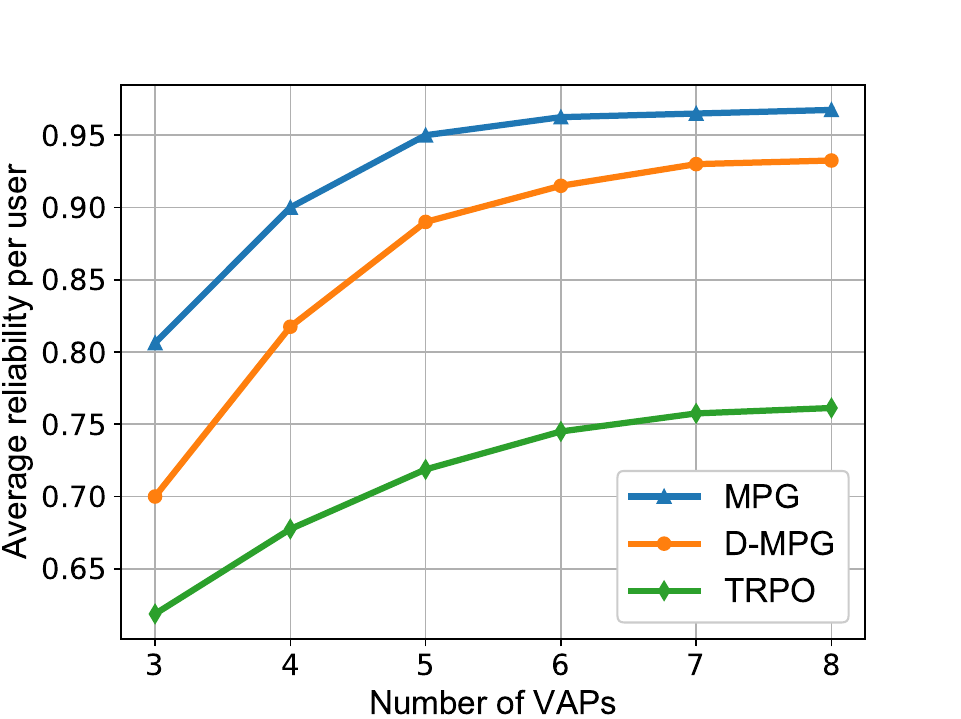}
%\captionsetup{labelfont={blue},textfont={blue}}
\caption{The average reliability per user as the number of SBSs varies ($B=7$).}
\label{fig12}
\end{figure}

Fig.~\ref{fig12} shows how the average reliability per user changes as the number of VAPs varies in a room with 7 SBSs. 
From Fig. \ref{fig12}, we can see that, as the number of VAPs increases, the average reliability per user of all algorithms increases.
The reason is that, as the number of VAPs increases, the central controller has more VAP options for localization so as to avoid blockages of VLC links.
Fig. \ref{fig12} also shows that the MPG and D-MPG algorithms can achieve up to 29.7\% and 21.2\% gains in terms of average reliability per user compared to the TRPO algorithm, respectively.
This is due to the fact the proposed algorithms can adapt to the dynamic users’ movement patterns and select the appropriate VAPs for user localization.

\begin{figure}[t]
\centering
\setlength{\belowcaptionskip}{-1cm}
\setlength{\abovecaptionskip}{0cm}
\includegraphics[width=8.7cm]{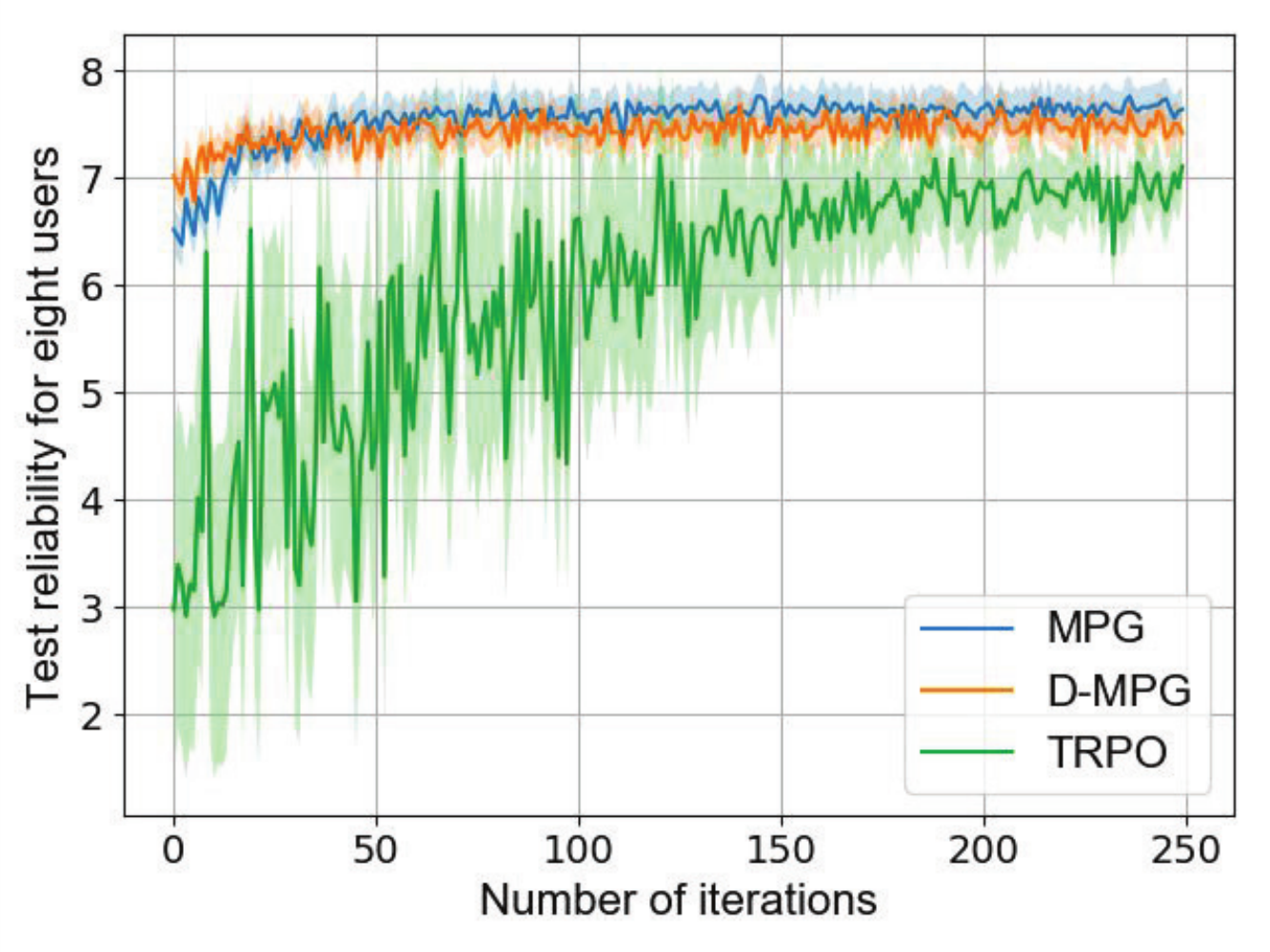}
%\captionsetup{labelfont={blue},textfont={blue}}
\caption{Test adaptability on new tasks.}
\label{fig8}
\end{figure}

Fig.~\ref{fig8} shows the adaptability of all the trained models testing for new tasks.
In Fig.~\ref{fig8}, the line and shadow are the mean and standard deviation computed over 5 random generated new tasks and all the models trained by old tasks are used as the initial models for the new tasks.
From Fig.~\ref{fig8}, we observe that, for new tasks, the proposed MPG and D-MPG algorithms achieve better performance at the beginning of the test process than the TRPO model.
This is because the policies trained by our proposed algorithms learned the knowledge useful for all tasks.
Fig.~\ref{fig8} also shows that the trained MPG and D-MPG models require approximately 30 and 20 iterations of further training to reach convergence for a new task, respectively, which are 81.2\% and 87.5\% less than the trained TRPO algorithm that requires about 160 iterations to reach convergence for a new task.
Meanwhile, compared with the training process in Fig.~\ref{fig4}, the test process of the MPG algorithm and D-MPG algorithm yield up to 88\% and 86.7\% reductions in terms of the number of iterations to reach convergence, respectively.
This demonstrates that the proposed MPG and D-MPG algorithms find the locally optimal policies that can quickly adapt to new tasks with new users' movement patterns.
In Fig.~\ref{fig8}, we can also see that the proposed MPG algorithm and D-MPG algorithm achieves up to 13.2\% and 10.3\% gains in terms of the reliability compared with the TRPO algorithm, respectively.
This is because the alternative iteration of the task learning step and the meta-learning step can find the broadly applicable parameters that can improve the performance of all tasks.

\begin{figure*}[t]
\vspace{-1cm}
\centering
\includegraphics[width=15.5cm]{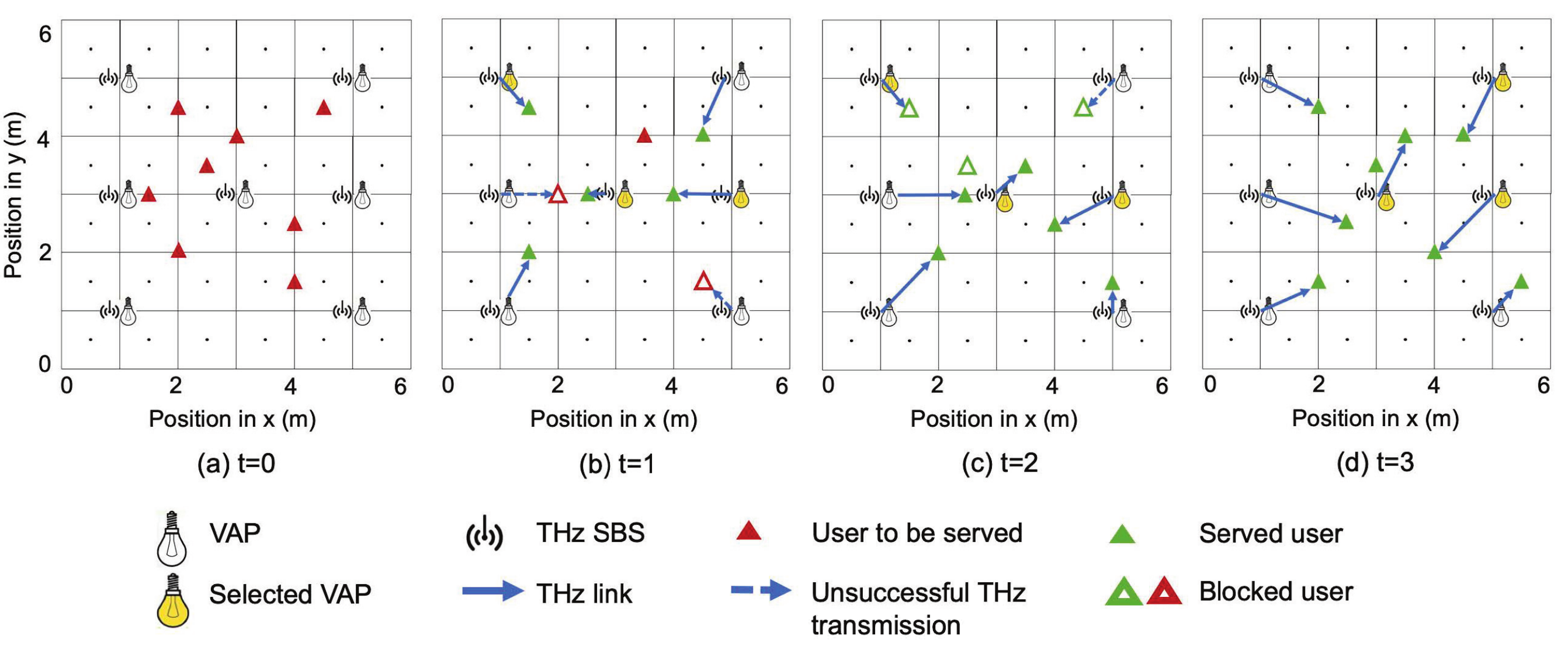}
\setlength{\abovecaptionskip}{0cm}
\caption{Visualization of using MPG algorithm for a VR scenario with eight users.}
\label{fig10}
\includegraphics[width=15.5cm]{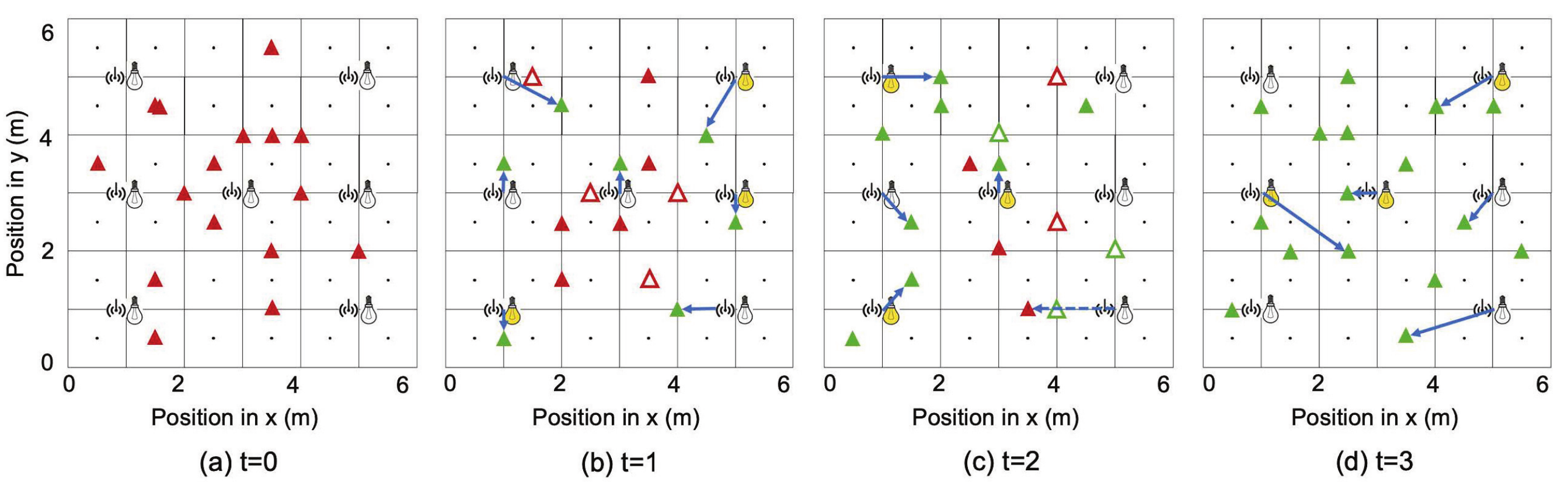}
\setlength{\belowcaptionskip}{-0.8cm}
\setlength{\abovecaptionskip}{0cm}
\caption{Visualization of using D-MPG algorithm for a VR scenario with sixteen users.}
\label{fig11}
\end{figure*}

In Figs.~\ref{fig10} and \ref{fig11}, we show examples of how the proposed MPG and D-MPG algorithms can optimize the reliability of the THz/VLC-enabled VR network, respectively.
In the examples, 7 VAPs and 7 SBSs are deployed in a $6$~m $\times$ $6$~m square room to serve VR users.
From Figs.~\ref{fig10} and \ref{fig11}, we can see that the VAP selection and the user association determined by the proposed algorithms can effectively serve the VR users during a time period and guarantee the network reliability.
This is because that the MPG and D-MPG algorithms have already learned the users' movement pattern during the sampling time period.
Moreover, the proposed algorithms aim to serve as many unserved users as possible so as to maximize the reward function.
Fig.~\ref{fig10} shows the VAP selection and the user-SBS association which result in the maximum reliability of randomly distributed 8 users.
In Figs.~\ref{fig10}(b) and \ref{fig10}(c) we can see that the unsuccessful THz transmission using the MPG algorithm is caused by the fact that the served user is not successfully localized and thus cannot establish a THz link.
Fig.~\ref{fig11} shows the VAP selection and the user-SBS association obtained by the D-MPG algorithm that can serve 16 users randomly distributed in the considered room.
From Fig.~\ref{fig11}(c), we can see that a user who cannot be localized blocks the THz link established according to the D-MPG algorithm.
In Fig.~\ref{fig11} we can also see that, compared with the user association determined by the MPG algorithm that all the SBSs need to serve users at each time slot as shown in Fig.~\ref{fig10}, the D-MPG algorithm only uses the necessary SBSs to provide transmission services to the unsuccessfully served users.
This is due to the fact that the D-MPG algorithm uses Hungarian algorithm to solve the dual problem of the user association optimization problem for those successfully localized users.

%The above results are obtained by running maml_trpo.py on HalfCheetahForwardBackwardEnv and AntForwardBackwardEnv for 300 updates. The figures show the expected sum of rewards over all tasks. The line and shadow are the mean and standard deviation computed over 3 random seeds.
\vspace{-0.2cm}
\section{Conclusion}
\vspace{-0.1cm}
\label{sec:6}
In this paper, we have developed a novel framework for maximizing reliability of THz/VLC-enabled wireless VR networks.  
To this end, we have formulated an optimization problem that jointly considers the user mobility, blockages of both THz and VLC links, VAP selection, and user association.
To solve this problem, we have developed a novel MPG algorithm based on meta-learning framework, which can effectively find the policy of VAP selection and user association for maximizing reliability.
The proposed MPG algorithm enables the trained policy to quickly adapt to new users' movement patterns.
Then, a D-MPG algorithm is proposed that use dual method to assist the MPG algorithm to determine user association so as to reduce the computational complexity of the MPG algorithm.
Simulation results have shown that, compared with the traditional RL algorithm, the proposed algorithms can achieve better performance and faster convergence speed. 
Simulation results also show that the proposed D-MPG algorithm can achieve the tradeoff between the algorithm processing time and the reliability gain.
In our future works, the use of massive MIMO and the handover overhead can be considered.
Meanwhile, the accuracy of the VLC-based indoor positioning and the deployment layout of VAPs and SBSs can be optimized to further improve the reliability of the studied VR network.
We can also consider the co-existence between VR users and cellular mobile users. 
\vspace{-0.2cm}


%\section*{Acknowledgment}




%\bibliographystyle{IEEEtran}

%\begin{thebibliography}{1}


%\end{thebibliography}
\bibliographystyle{IEEEbib}
\def\baselinestretch{1}
\bibliography{VLC_THz}

\end{document}